\documentclass[12pt]{article}

\usepackage[height=8.85in,width=6.45in]{geometry}
\usepackage{ amsmath,amssymb,amsfonts,amsxtra, mathrsfs, makeidx,graphics,graphicx,amsthm,epsfig, ytableau,bm,longtable,float, color,tikz,mathtools,xfrac,footnote,rotating, lscape, makecell, environ,mathtools, empheq}
\usepackage{hyperref}

\usepackage{times}
\usepackage{courier}
\usepackage{braket}
\usepackage{mathtools}
\numberwithin{equation}{section}
\usepackage{xcolor}

\usepackage{mdframed}

\def\Nequals#1{$\mathcal{N}{=}\,#1$}
\def\bC{\mathbb{C}}
\def\bH{\mathbb{H}}
\def\bR{\mathbb{R}}
\def\bZ{\mathbb{Z}}
\def\cM{\mathcal{M}}
\def\cO{\mathcal{O}}
\def\cT{\mathcal{T}}
\def\fg{\mathfrak{g}}
\def\fe{\mathfrak{e}}
\def\fsu{\mathfrak{su}}
\def\fso{\mathfrak{so}}
\def\fusp{\mathfrak{usp}}
\def\tr{\mathop{\mathrm{tr}}}
\def\SO{\mathrm{SO}}
\def\SU{\mathrm{SU}}
\def\USp{\mathrm{USp}}
\def\UU{\mathrm{U}}
\def\E#1#2#3#4#5#6#7#8#9{%
\begin{array}{c@{\,}c@{\,}c@{\,}c@{\,}c@{\,}c@{\,}c@{\,}c}
&&&&&#9\\
#1&#2&#3&#4&#5&#6&#7&#8
\end{array}}
\def\un#1{\underline{#1}}
\def\vev#1{\langle#1\rangle}
\def\hkq{/\!/\!/}

\usetikzlibrary{positioning}
\usetikzlibrary{chains}
\usetikzlibrary{trees}
\usetikzlibrary{arrows,fit,decorations.pathreplacing}
\tikzstyle{every picture}+=[remember picture]
\tikzstyle{na} = [baseline]

\usetikzlibrary{arrows, decorations.markings, calc, fadings, decorations.pathreplacing, patterns, decorations.pathmorphing, positioning}

\def\node#1#2{\overset{#1}{\underset{#2}{{\color{gray} \bullet}}}}
\def\Node#1#2{\overset{#1}{\underset{#2}{{ \bullet}}}}

\def\sqnode#1#2{\overset{#1}{\underset{#2}{{\color{gray} \blacksquare}}}}
\def\sqNode#1#2{\overset{#1}{\underset{#2}{{ \blacksquare}}}}
\def\cver#1#2{\overset{{\llap{$\scriptstyle#1$}\displaystyle{\color{gray} \bullet}{\rlap{$\scriptstyle#2$}}}}{\scriptstyle\vert}}

\def\uer#1#2{\underset{{\llap{$\scriptstyle#1$}\displaystyle{\color{gray} \bullet}{\rlap{$\scriptstyle#2$}}}}{\scriptstyle\vert}}

\newcommand{\be}{\begin{equation}} \newcommand{\ee}{\end{equation}}
\newcommand{\bea}{\begin{equation} \begin{aligned}} \newcommand{\eea}{\end{aligned} \end{equation}}
\newcommand{\ba}{\begin{array}}
\newcommand{\ea}{\end{array}}
\newcommand{\bi}{\begin{itemize}}
\newcommand{\ei}{\end{itemize}}
\def\vec#1{\bm{#1}}
\def\bea#1\eea{\allowdisplaybreaks \begin{align}#1\end{align}}
 \newcommand{\ben}{\begin{enumerate}}
\newcommand{\een}{\end{enumerate}}
\newcommand{\bean}{\begin{eqnarray*}}
\newcommand{\eean}{\end{eqnarray*}}
\newcommand{\eref}[1]{(\ref{#1})}

\def\CM{{\cal M}}
\def\CN{{\cal N}}

\def\CS{{\cal S}}
\def\CT{{\cal T}}

\newcommand{\BC}{\mathbb{C}}

\newcommand{\BH}{\mathbb{H}}

\newcommand{\nT}{N_S}
\newcommand{\NN}{N_3}
\newcommand{\Nprime}{N_\text{inst}}
\renewcommand{\kappa}{{N_6}}

\begin{document}

\begin{titlepage}

\begin{flushright}
IPMU-17-0097
\end{flushright}

\vskip 4cm

\begin{center}

{\Large\bfseries $E_8$ instantons on type-A ALE spaces \\[.2em]
and supersymmetric field theories  }

\vskip 1cm
Noppadol Mekareeya$^1$, Kantaro Ohmori$^2$, 
Yuji Tachikawa$^3$
and Gabi Zafrir$^3$
\vskip 1cm

\begin{tabular}{ll}
1 & Dipartimento di Fisica, Universit\`a di Milano-Bicocca, and\\
& INFN, sezione di Milano-Bicocca, \\
& Piazza della Scienza 3, I-20126 Milano, Italy \\
2 & School of Natural Sciences, Institute for Advanced Study, \\
& Princeton, NJ 08540, USA\\
3  & Kavli Institute for the Physics and Mathematics of the Universe, \\
& University of Tokyo,  Kashiwa, Chiba 277-8583, Japan\\
\end{tabular}

\vskip 1cm

\end{center}

\noindent 
We consider the 6d superconformal field theory realized on M5-branes probing the $E_8$ end-of-the-world brane on the deformed and resolved $\bC^2/\bZ_k$ singularity.
We give an explicit algorithm which determines, for arbitrary holonomy at infinity,
the 6d quiver gauge theory on the tensor branch,
the type-A class S description of the $T^2$ compactification, and
the star-shaped quiver obtained as the mirror of the $T^3$ compactification.

\end{titlepage}

\tableofcontents

\section{Introduction and summary}

One of the many surprises during the second superstring revolution was the realization that the  construction of  $\SU(N)$ instantons  on $\bR^4$ by Aityah-Drinfeld-Hitchin-Manin \cite{Atiyah:1978ri} and on asymptotically locally Euclidean (ALE) spaces by Kronheimer-Nakajima \cite{KronheimerNakajima,Bianchi:1996zj} have a physical realization in terms of D$p$-branes probing D$(p{+}4)$-brane on $\bR^4$ \cite{Witten:1995gx} and/or ALE spaces \cite{Douglas:1996sw}.
There, we have a gauge theory with eight supercharges on D$p$-branes such that  its Higgs branch is given by the corresponding instanton moduli spaces: the equations of the ADHM and Kronheimer-Nakajima construction are the F-term and D-term conditions of the supersymmetric gauge theory.

As a variation of this construction, we can consider M5-branes probing the $E_8$ end-of-the-world brane of the M-theory, either on $\bR^4$ or on ALE spaces.
The low-energy worldvolume theory on these M5-branes is a 6d \Nequals1 supersymmetric theory whose Higgs branch is intimately related to the $E_8$ instanton moduli spaces.
These theories were studied already in the heyday of the second revolution, see e.g.~\cite{Aspinwall:1997ye},
We did not, however, have many methods to understand the properties of these theories back then, 
since these theories are intrinsically strongly-coupled.
Therefore we could not say anything new regarding the mathematics of the $E_8$ instanton moduli spaces on $\bR^4$ or ALE spaces, for which no constructions analogous to ADHM or Kronheimer-Nakajima are known even today.

The situation has changed drastically since then, thanks to our improved understanding of strongly-coupled supersymmetric theories. 
Among others, we can count the class S construction in four dimensions initiated by \cite{Gaiotto:2009we},
the determination of the chiral ring of the Coulomb branch in three dimensions starting with \cite{Cremonesi:2013lqa},
and a new method to study \Nequals1 theories in six dimensions pioneered by \cite{Heckman:2013pva}.
Combining these developments, we believe there might be a chance that the physics might shed new lights on the mathematics of the structure of the $E_8$ instanton moduli spaces on ALE spaces.

The main target of our study in this paper is the 6d \Nequals1 theory on M5-branes probing the $E_8$ 9-brane on the A-type ALE singularity $\bC^2/\bZ_k$.
Such a system can be labeled by the M5-brane charge $Q$ and the asymptotic holonomy $\rho:\bZ_k \to E_8$.
For some simple choices of $\rho$, the structure of the 6d theory on generic points on its tensor branch was already determined in \cite{DelZotto:2014hpa,Heckman:2015bfa}, which was further extended in \cite{Zafrir:2015rga,Ohmori:2015tka,Hayashi:2015zka}.

Our first aim is to determine the tensor branch structure for an \emph{arbitrary} choice of the asymptotic holonomy $\rho$.
We give a complete algorithm determining the gauge group and the matter content in terms of $\rho$.
Along the way, we encounter a subtle feature that there are two distinct ways to gauge $\fsu(2n+8)$ symmetry of $\fso(4n+16)$ flavor symmetry of an $\fusp(2n)$ gauge theory with $N_f=2n+8$ flavors, due to the fact that the outer automorphism of $\fso(4n+16)$ is \emph{not} a symmetry of the latter gauge theory.

The Higgs branch  $\cM_{Q,\rho}$ of our theory  $\cT_{Q,\rho}^\text{6d}$ is not directly the instanton moduli space.
In particular, $\cM_{Q,\rho}$ has an action of $\SU(k)$, which we do not expect for the instanton moduli space.
Rather, by a small generalization of the argument in \cite{Ohmori:2015pua}, we see that the $E_8$ instanton moduli space $\cM^\text{inst}_{Q,\rho,\xi}$ of charge $Q$ and asymptotic holonomy $\rho$ on the ALE space $\widetilde{\bC^2/\bZ_k}$ is given by \begin{equation}
\cM^\text{inst}_{Q,\rho,\xi} = (\cM_{Q,\rho} \times \cO_\xi) \hkq \SU(k) \label{rel}
\end{equation}
where $\xi=(\xi_\bC,\xi_\bR)\in\fsu(k)\otimes (\bC\oplus \bR)$ is an element in the Cartan of $\fsu(k)$ tensored by $\bR^3$ specifying the hyperk\"ahler deformation parameter of the ALE space,
$\cO_\xi$ is the orbit of $\xi_\bC$ in $\fsu(k)_\bC$ with the hyperk\"ahler metric specified by $\xi_\bR$ as in \cite{KronheimerNilpotent},
and the symbol $\hkq$ denotes the hyperk\"ahler quotient construction.
This means that the space $\cM_{Q,\rho}$ knows the structure of the instanton moduli on the ALE space for arbitrary deformation parameter $\xi$.
The existence of such a generating space was conjectured by one of the authors in \cite{Tachikawa:2014qaa}, based on a study of $\SO(8)$ instantons on the ALE spaces.

We then study the 4d theory which arises from the $T^2$ compactification of the 6d theory as in \cite{Ohmori:2015pua}. 
We find that they always correspond to a class S theory of type A, given by a sphere with three punctures.
The 3d mirror of its $S^1$ compactification is a star-shaped quiver, whose structure can be deduced from the class S description by the methods of \cite{Benini:2010uu}.
We find that they have the form of an \emph{over-extended} $E_8$ quiver. 
In 3d, the relation \eqref{rel} can be physically implemented by realizing $\cO_\xi$ as the Coulomb branch of the $T[\SU(k)]$ theory.
Using this, we will find that $\cM^\text{inst}_{Q,\rho,\xi}$ is the Higgs branch of an affine $E_8$ quiver where $\xi$ is now the mass parameter of an $\SU(k)$ flavor symmetry.
For $\xi=0$ this was already conjectured by mathematicians \cite{Nakajima:2015txa,Nakajima:2015gxa} and by physicists \cite{Cremonesi:2014xha,Mekareeya:2015bla}.

\paragraph{Organization of the paper:}
The rest of the paper is organized as follows.
We start by recalling the geometric data characterizing our system in Sec.~\ref{sec:geometric}.
Then in Sec.~\ref{sec:6d}, we provide the algorithm determining the 6d quiver theory in terms of the asymptotic holonomy.
In Sec.~\ref{sec:lower}, we discuss its dimensional reduction to 5d, 4d and 3d in turn.
In 5d and 4d, we translate the Kac labels to the three Young diagrams characterizing the brane web and the class S description. 
In 3d, we give the star-shaped quiver.
Finally in Sec.~\ref{sec:examples}, we provide many examples illustrating our discussions.

\paragraph{Accompanying Mathematica file:}
The paper comes with a Mathematica file which implements the algorithm to produce the 6d quiver given the asymptotic $E_8$ holonomy. In addition, it allows the user to determine the 4d class S theory, and compute the anomalies from three different methods, namely the 6d field theory, the M-theoretic inflow, and the 4d class S technique.

\paragraph{Summary of Notations:}
\begin{itemize}
\item The asymptotic holonomy $\rho:\bZ_k\to E_8$ is given by an element $\vec w\in \fe_8$ in the Cartan subalgebra, or equivalently in terms of the Kac label \begin{equation}
\un{n}:=\E{n_1}{n_2}{n_3}{n_4}{n_5}{n_6}{n_{4'}}{n_{2'}}{n_{3'}},
\end{equation} a set of non-negative integers arranged on the affine $E_8$ Dynkin diagram.
For more details, see Sec.~\ref{sec2.1}.
\item We have closely related quantities $\Nprime $, $\NN$, $\nT $, $\kappa$, and $Q$, which are all essentially the number of M5-branes or equivalently the instanton charge on the ALE space. 
They all increase by one when we add one M5-brane to the system.
Their constant parts are however different.
We could have used just one out of them, but any choice would make at least one of the formulas quite unseemly. 
We therefore decided to keep them and provide a summary here.
\begin{itemize}
\item The integer $\Nprime $ is defined in terms of the instanton number as \begin{equation}
\int_{\widetilde{\bC^2/\Gamma}}\tr F\wedge F \propto \Nprime -\frac{\vev{\vec w,\vec w}}{2k}.
\end{equation}
See \eqref{E8instantonnumber} for details.
\item Another integer $\NN$ satisfies $\NN=\Nprime -k$, see \eqref{NprimeN0}, \eqref{NprimeN}. This  is useful to parameterize the ranks of groups in the 3d quiver, see \eqref{N3d}.
\item Another integer $\nT$ defined by  $\nT =\NN+n_1+\cdots+n_6$ is  useful to parameterize the class S data, see \eqref{N}.
\item The integer $\kappa$ is the number of tensors of the 6d quiver.  The difference between $\nT $ and $\kappa$ is determined by the Kac label and is described in the algorithm in Sec.~\ref{algorithm}.
\item A rational number $Q$ is the M5-charge which appears in the inflow computation, and satisfies
\begin{equation}
Q=\Nprime -\frac{\vev{\vec w,\vec w}}{2k}-\frac12(k-\frac1k),
\end{equation}see \eqref{M5charge}.
\end{itemize}
\end{itemize}

\section{Geometric preliminaries}
\label{sec:geometric}
\subsection{Topological data of the instanton configuration}
\label{sec2.1}
Here we recall the topological data necessary to specify a $G$-instanton on $\bC^2/\Gamma$ or its resolution $\widetilde{\bC^2/\Gamma}$, where $\Gamma\in \SU(2)$.

On $\bC^2/\Gamma$, we first need to specify the holonomy at the origin and at the infinity. 
They determine the representation $\rho_{0,\infty}: \Gamma\to G$, which we consider as a linear action on the complexified adjoint representation $\fg$. 

On $\widetilde{\bC^2/\Gamma}$, we specify the holonomy at infinity $\rho_\infty$.
In addition, we need to specify the class in $H^2(\widetilde{\bC^2/\Gamma},\pi_1(G))$.
This is the first Chern class when $G=\UU(N)$ and the second Stiefel-Whitney class when $G=\SO(N)$.

Finally we need to specify the instanton number, defined as the integral of $\tr F\wedge F$ over the ALE space. 
Unless otherwise mentioned, we normalize the trace  so that the instanton on $\bR^4$ of the smallest positive instanton number satisfies \begin{equation}
\int \tr F\wedge F=1.
\end{equation}
On the ALE space, the instanton number is in general fractional.

Our main interest lies in the case $G=E_8$ and $\Gamma=\bZ_k$.
Since $\pi_1(E_8)$ is trivial, we do not have to specify the class in $H^2$.

A holonomy $\rho:\bZ_k \to E_8$ can be nicely encoded by its Kac label
\begin{equation}
\un{n}:=\E{n_1}{n_2}{n_3}{n_4}{n_5}{n_6}{n_{4'}}{n_{2'}}{n_{3'}}.
\end{equation}
 introduced in \S 8.6 of Kac's textbook \cite{Kac}. 
Let us quickly recall how it works. 
Let the image $g$ of the generator of $\bZ_k$ in $E_8$ be \begin{equation}
g=e^{2\pi i \vec w/k}\in E_8
\end{equation} where \begin{equation}
\vec w=\sum_{i\neq 0} n_i \vec w_i \in \fe_8.
\end{equation}
where $\vec w_i$ are the fundamental weights of $E_8$. 
Since $g$ is of order $k$, $n_i$ are integers. 
We define $n_0$ so that  $\sum d_i n_i = k$, where  the Dynkin marks $\un{d}$ are given by \begin{equation}
\un{d}=\E123456423.
\end{equation}
It is known that by the Weyl reflections and the shifts, 
we can arrange $n_i\ge 0$ for all $i$ and then the result is unique. 
This is the Kac label of the holonomy. 

The subalgebra of $\fe_8$ left unbroken by the holonomy $\rho$ can be easily read off from its Kac label.
Namely, it is given by the subalgebra corresponding to the nodes $i$ of the Dynkin diagram where $n_i=0$, together with an Abelian subalgebra making the total rank $8$.

On $\widetilde{\bC^2/\bZ_k}$, 
the instanton number modulo one is given by the classical Chern-Simons invariant evaluated on $S^3/\bZ_k$ at infinity. 
One way to compute it is to introduce coordinates on $S^3/\bZ_k$ using polar coordinates $\theta,\phi$ on $S^2$ and the angle $\psi$ along the $S^1$ fiber. 
The connection itself is $\propto \vec w (d\psi + \cdots)$. 
One finds that \begin{equation}
\int\tr F\wedge F = \Nprime - \frac{\vev{\vec w, \vec w}}{2k}\label{E8instantonnumber}
\end{equation} 
where $\Nprime $ is an integer.
The hyperk\"ahler dimension of the moduli space is given by the formula 
 \begin{equation}
\dim_\bH \mathcal{M}_{\widetilde{\bC^2/\Gamma},\rho_\infty} =  30 \Nprime  - \vev{\vec w,\vec \rho}.
\label{geometry-result}
\end{equation}

\subsection{Dimension of the instanton moduli space}
In this subsection we derive the formula \eqref{geometry-result} of the dimension of the moduli space.
Those readers who trust the authors can skip this subsection.
This computation is of course not new. 
It is provided here to make this paper more self-contained.

The basic tool is the Atiyah-Patodi-Singer index theorem.
Its explicit form on the orbifold of $\bC^2$ was worked out e.g.~in \cite{Nakajima} for $\Gamma\subset \mathrm{U}(2)$.
Here we quote the form used in Kronheimer-Nakajima \cite{KronheimerNakajima} for $\Gamma\subset\mathrm{SU}(2)$.  
The formula for the orbifold is:
\begin{multline}
\dim_\bH \mathcal{M}_{\bC^2/\Gamma,\rho_\infty,\rho_0} = 
h^\vee(G) (\int\tr F\wedge F )   +  \frac{1}{2|\Gamma|}\sum_{\gamma \neq e} \frac{\chi_{\rho_\infty}(\gamma) }{2-\chi_Q(\gamma)}
-  \frac{1}{2|\Gamma|}\sum_{\gamma \neq e} \frac{\chi_{\rho_0}(\gamma) }{2-\chi_Q(\gamma)}.
\end{multline}
Here,
$h^\vee(G)$ is the dual Coxeter number of $G$,
and the second and the third terms are the contributions from the $\eta$ invariant of $S^3/\Gamma$ at the asymptotic infinity and at the origin, respectively,
and  $Q$ is the standard two-dimensional representation of $\Gamma$ from the defining embedding $\Gamma\subset \SU(2)$,

On the ALE space $\widetilde{\bC^2/\Gamma}$, we have: \begin{equation}
\dim_\bH \mathcal{M}_{\widetilde{\bC^2/\Gamma},\rho_\infty} = 
h^\vee(G) (\int\tr F\wedge F )   +  \frac{1}{2|\Gamma|}\sum_{\gamma \neq e} \frac{\chi_{\rho_\infty}(\gamma) }{2-\chi_Q(\gamma)}
-  \frac1{24}\dim G \chi_\Gamma 
\end{equation}
where the quantity
\begin{equation}
\chi_\Gamma:=r_\Gamma +1 -\frac{1}{|\Gamma|} \label{chiGamma}
\end{equation}
 is the Euler number of $\widetilde{\bC^2/\Gamma}$ as defined by the integral of the Pontrjagin density.
Furthermore,  \begin{equation}
\frac1{24} (r_\Gamma +1 -\frac{1}{|\Gamma|}) = \frac{1}{2|\Gamma|}\sum_{\gamma \neq e} \frac{1 }{2-\chi_Q(\gamma)},
\end{equation} reflecting the fact  that if the holonomy at the origin of an instanton on $\bC^2/\Gamma$ is trivial, we can resolve/deform the instanton and the ALE at the same time to be on $\widetilde{\bC^2/\Gamma}$. 
In the end, we find the formula \begin{equation}
\dim_\bH \mathcal{M}_{\widetilde{\bC^2/\Gamma},\rho_\infty} = 
h^\vee(G) (\int\tr F\wedge F )   +  \Delta\eta, 
\label{intermediate}
\end{equation}
where
\begin{equation}
 \Delta\eta:=\frac{1}{2|\Gamma|}\sum_{\gamma \neq e} \frac{(\chi_{\rho_\infty}(\gamma) -\dim \fg) }{2-\chi_Q(\gamma)}
\end{equation}

Let us evaluate this formula when $G=E_8$ with the holonomy $\rho_\infty$ specified by $g=e^{2\pi i \vec w/k}$ with the Kac label $\un{n}$.
The eta invariant is \begin{equation}
\Delta\eta =\frac{1}{2|\Gamma|}\sum_{\gamma \neq e} \frac{(\chi_{\rho_\infty}(\gamma) -\dim \fg) }{2-\chi_Q(\gamma)} = \frac{1}{2k} \sum_{\vec \alpha:\text{all roots}} \sum_{\gamma\neq 1}\frac{\chi_{\vev{\vec w,\vec \alpha}}(\gamma)-1 }{2-\chi_Q(\gamma)}
\end{equation}
where \begin{equation}
\chi_a(g^j) = e^{2\pi i aj/k}, \qquad \chi_Q(g^j) = 2\cos 2\pi j/k.
\end{equation} Now, we note \begin{equation}
\frac{1}{2k}\sum_{j=1}^{k-1} \frac{\chi_a(g^j) -1}{2-\chi_Q(g^j)} = - \frac{a(a-k)}{4k}
\end{equation} for $a=0,1,\ldots k$.
Then \begin{equation}
\Delta\eta = 2\sum_{\vec\alpha:\text{positive roots}} -\frac{\vev{\vec \alpha,\vec w} (\vev{\vec \alpha,\vec w}-k) }{4k}
\end{equation} since $0\le \vev{\vec\alpha,\vec w}\le k$ for positive roots $\vec\alpha$.
Now we use \begin{equation}
\sum_{\vec\alpha:\text{positive roots}} \vec\alpha = 2\rho,\qquad
\sum_{\vec\alpha:\text{positive roots}} \vev{\vec v_1,\vec\alpha}\vev{\vec\alpha,\vec v_2} = h^\vee \vev{\vec v_1,\vec v_2}
\end{equation} and find \begin{equation}
\Delta\eta=\frac{h^\vee}{2k} \vev{\vec w,\vec w} - \vev{\vec w,\vec \rho}.
\end{equation}

To compute the dimension, we now plug in to \eqref{intermediate} the formula for $\Delta\eta$ found just above and the formula for the instanton number \eqref{E8instantonnumber}.
The term proportional to $\vev{\vec w,\vec w}$ cancels out, and we indeed have the desired result \eqref{geometry-result}.

\section{Six-dimensional description}
\label{sec:6d}
After these geometrical preliminaries, we move on to the field theoretical analysis.
We start with the six-dimensional quiver descriptions. 
As already mentioned in the introduction, for various simple choices of $\rho$, the six-dimensional quivers were already determined in \cite{DelZotto:2014hpa,Heckman:2015bfa,Zafrir:2015rga,Ohmori:2015tka,Hayashi:2015zka}.
By a series of trials and errors, and following the principle that the quiver should be  determined in terms of the Kac label, the authors found the following algorithm. 

\subsection{The general structure of the quiver}

Our 6d SCFT on the generic points on its tensor branch consists of a collection of $\kappa$ tensors,
corresponding to a linear quiver of the form 
\begin{equation}
G_1\times \SU(m_2) \times \SU(m_3)\times \cdots \times \SU(m_\kappa)\times [\SU(k)]
\end{equation}
where $G_1$ is on the $-1$ curve, the rest is on $-2$ curves, and the final $\SU(k)$ is a flavor symmetry.
In the notation of \cite{Heckman:2015bfa}, we have
\begin{equation}
\overset{G_1}{1} \quad
\overset{\fsu(m_2)}{2} \quad
\overset{\fsu(m_3)}{2} \quad
\cdots\quad
\overset{\fsu(m_\kappa)}{2} \quad
[\SU(k)].
\end{equation}

Below, we slightly abuse the notation and refer by $G_1$ the combination of the group and the non-fundamental hypermultiplets on the $-1$ curve. 
The choices are:
\begin{itemize}
\item $G_1=\USp(m_1)$,
\item $G_1=\SU(m_1)$ with an antisymmetric hyper, or 
\item $G_1=\SU(m_1=6)$ with a rank $3$ antisymmetric half-hyper. 
\end{itemize}
We consider the rank $1$ E-string theory as $\USp(0)$, and furthermore,
the rank $2$ E-string theory is considered as a $\USp(0)$ connecting to an $\SU(1)$ group.

We have $m_1\le m_2\le \cdots \le m_\kappa$,
and we define $a_1,\ldots, a_9$ by \begin{equation}
a_s = \#\{i \mid m_{i+1}-m_i = s\}.
\end{equation}
We can reconstruct the whole of $m_i$ from $m_1$, $\kappa$ and $a_1,\ldots a_9$.
For example, when the quiver is 
\begin{equation}
\SU(3) \times \SU(9) \times \SU(13) \times \SU(17) \times \SU(18) \times \SU(19) \times \SU(20)
\end{equation}
we have $a_8=a_7=a_5=a_3=a_2=0, a_6=1, a_4=2, a_1=3$. 

There are  bifundamentals between two consecutive groups in the quiver,
and finally fundamental hypers are added such that each group is anomaly free, that is:
\begin{itemize}
\item $N_f = 2N$ for $\SU(N)$,
\item $N_f = N +8$ for $\USp(N)$ or $\SU(N)$ with an antisymmetric hyper, and
\item $N_f=15$ for $\SU(6)$ with a rank $3$ antisymmetric half-hyper.
\end{itemize}

\subsection{The algorithm}
\label{algorithm}

Now we present the algorithm to determine the structure of the quiver given the Kac label $\un{n}$ and the number $\kappa$ of the groups. 
Along the way, we also define the quantity $\nT $ which will be used in the following.
We will also need the quantity \begin{equation}
\NN=\nT -n_1-n_2-n_3-n_4-n_5-n_6.\label{N}
\end{equation}
The algorithm is implemented in the accompanying Mathematica file, so that the reader can easily try it around.

In general we have \begin{equation}
a_i=n_i\quad\text{for}\quad i=1,2,3,4,5,6.
\end{equation}
To specify $a_{7,8,9}$, we need to consider various cases as summarized below:
\[
\begin{cases}
n'_4 \ge n'_3 & 
	\longrightarrow\begin{cases}
	n'_4-n'_3=\text{even} & \longrightarrow\text{Case 1}, \\
	n'_4-n'_3=\text{odd} & \longrightarrow\text{Case 2}, 
	\end{cases}\\[2em]
n'_3 \ge n'_4 & 
	\longrightarrow\begin{cases}
	n'_2 < (n'_3-n'_4)/2 & \longrightarrow \text{Case 4}, \\	
	n'_2 \ge (n'_3-n'_4)/2 & \longrightarrow\begin{cases}
		n'_3-n'_4=\text{odd} & \longrightarrow\text{Case 3},\\
		n'_3-n'_4=\text{even}& \longrightarrow\text{Case 5}.
		\end{cases}
	\end{cases}
\end{cases}
\]
For each case, the output of the algorithm is $(a_{7,8,9}, G_1, \nT)$ as shown below:
\begin{enumerate}
\item $n'_4 \geq n'_3, n'_4 - n'_3 = \text{even}$: 
	\begin{itemize}
	\item $a_7 = n'_3, a_8 = \frac{n'_4 - n'_3}{2}, a_9 = 0$. 
	\item $G_1=\USp(2 n'_2)$. 
	\item $\nT  = \kappa-\frac{n'_4 + n'_3}{2}$.
	\end{itemize}
\item $n'_4 \geq n'_3 + 1, n'_4 - n'_3 = \text{odd}$: 
	\begin{itemize}
	\item $a_7 = n'_3, a_8 = \frac{n'_4 - n'_3 - 1}{2}, a_9 = 0$. 
	\item $G_1=\SU(2 n'_2 +4)$ group with an antisymmetric hyper. 
	\item $\nT  = \kappa-\frac{n'_4 + n'_3 - 1}{2}$.
	\end{itemize}
\item $n'_3 \geq n'_4 + 1, n'_3 - n'_4 = \text{odd}, n'_2 \geq \frac{n'_3 - n'_4 - 1}{2}$: 
	\begin{itemize}
	\item $a_7 = n'_4, a_8 = \frac{n'_3 - n'_4 - 1}{2}, a_9 = 0$. 
	\item $G_1=\SU(2 n'_2 + n'_4 - n'_3 + 4)$ group with an antisymmetric hyper. 
	\item $\nT  = \kappa-\frac{n'_4 + n'_3 - 1}{2}$.         
	\end{itemize}
\item $n'_3 > n'_4 + 2 n'_2 + \ell, n'_3 - n'_4 - 2 n'_2 = 3 x + \ell, x \in \bZ, \ell=0,1,2$: 
	\begin{itemize}
	\item $a_7 = n'_4, a_8 = n'_2, a_9 = \frac{n'_3 - n'_4 - 2 n'_2 - \ell}{3}$. 
	\item $G_1$ is 
		\begin{itemize}
		\item Rank $1$ $E_8$ for $\ell=0$,
		\item  $\SU(3)$ for $\ell=1$,
		\item $\SU(6)$ with a half-hyper in the rank $3$ antisymmetric for $\ell=2$. 
		\end{itemize}
	\item $\nT  = \kappa-\frac{n'_3 + 2 n'_4 + n'_2 - l}{3}$.
	\end{itemize}
\item $n'_3 \geq n'_4,  n'_3 - n'_4 = \text{even}, n'_2 \geq \frac{n'_3 - n'_4}{2}$: 
	\begin{itemize}
	\item $a_7 = n'_4, a_8 = \frac{n'_3 - n'_4}{2}, a_9 = 0$. 
	\item $G_1=\USp(2 n'_2 + n'_4 - n'_3)$. 
	\item $\nT  = \kappa-\frac{n'_4 + n'_3}{2}$.
	\end{itemize}
\end{enumerate}

\subsection{A subtlety concerning the 6d $\theta$ angle}
\label{sec:subtlety}
Note that the quivers produced in Case 5 are the same ones as the ones produced by Case 1,
as far as the data we described so far are concerned.
This is perfectly fine when $n'_4=n'_3$, since in this case we are just applying the different cases to the same Kac label.
However, when $n'_4\neq n'_3$, or equivalently when $a_8\neq 0$,
the resulting quivers should however be subtly different, since e.g.~they reduce to different 4d class S theories and 3d star-shaped quivers.
We argue that the difference between them is how one embeds the $\SU(2N+8)$ group into the $\SO(4N+16)$ global symmetry group of $\USp(2N)$. 

A relatively simple case is the following.
Let us first consider the cases when $n'_4 = 2, n'_3 = 0 , n'_2 = 0$ versus $n'_4 = 0, n'_3 = 2, n'_2 = 1$, with the rest of labels being zero $n_{1,\ldots,6}=0$. 
Both theories have the form of a long $\SU(8)$ quiver gauging an $\SU(8)$ subgroup of the rank $1$ $E_8$ theory. The two differ by the embedding of $\SU(8)$ inside $E_8$ and in fact have different global symmetries. 
To see this, consider embedding $\SU(8)$ inside $\SO(16) \subset E_8$. The adjoint of $E_8$ decomposes under its $\SO(16)$ maximal subgroup as $\bold{248}\rightarrow \bold{120} + \bold{128}$. 
Now consider decomposing $\SO(16)$ to its $\UU(1)\times \SU(8)$ maximal subgroup. 
Under this embedding the spinors of $\SO(16)$ decompose to the rank $x$ antisymmetric tensors of $\SU(8)$ for $x=0,2,4,6,8$ for one spinor and $x=1,3,5,7$ for the other. 
However only one spinor appears in the adjoint of $E_8$, and therefore there are two different embedding of $\SU(8)$ inside $E_8$. 
In one of them the $\bold{128}$ contains gauge invariant contributions leading to the larger global symmetry.

The general case corresponds to the situation where $\SU(2N+8)$ is embedded in $\SO(4N+16)$.
There is no distinction in the perturbative sector of the theory. However the theory possesses instanton strings. 
The ones for $\USp$ groups will be in a chiral spinor of the $\SO$ group and so will decompose differently depending on the embedding. 
This then leads to theories with distinct spectrum of string excitations. 
Also note that this only occurs if the entire $\SO$ symmetry is gauged leaving only a $\UU(1)$ commutant. 
If we gauge an $\SU(x) \subset \SO(2x)\subset \SO(4N+16)$ with $x<2N+8$,
then the chiral spinor of $\SO(4N+16)$ decomposes to non-chiral spinors of $\SO(2x)$ and therefore there is a single embedding. 
This agrees with the fact that the cases coincide when $a_8=0$. 

We can understand this distinction from the existence of the discrete $\theta$ angle in 6d, due to the fact that $\pi_5(\USp(2N))_5 = \bZ_2$.
Suppose now that the $\USp$ group has $2n$ half-hypermultiplets in the fundamental.
Classically it has an $\mathrm{O}(2n)$ flavor symmetry,
but the parity part flips the discrete theta angle. 
Therefore the flavor symmetry is actually $\fso(2n)$.
The two embeddings of $\fsu(n)$ into $\fso(2n)$ are related exactly by the parity part of $\mathrm{O}(2n)$, and therefore are inequivalent. 
The F-theoretical interpretation of these two inequivalent embeddings seems to be unknown.
It would be interesting to work it out.\footnote{The authors thank D. R. Morrison for the correspondence on this point.}

Note that  an analogous phenomenon exists in $5d$, where given a pure $\USp$ group there are two distinct $5d$ SCFTs associated with this theory differing by the instanton spectrum of the $5d$ gauge theory.
This is related to the existence of a $\bZ_2$ valued $\theta$ angle originating from the fact that $\pi_4(\USp(2N))_4 = \bZ_2$.

\subsection{Anomalies and the inflow}
The anomaly of these 6d SCFTs can be computed from their quiver description using the technique of \cite{Ohmori:2014kda,Intriligator:2014eaa}.
We should be able to match it to the anomaly computed from the inflow using the M-theory description.

The inflow computations of M5-branes probing the $E_8$ end-of-the-world brane
and of M5-branes probing the $\bC^2/\bZ_k$ singularity
was given in \cite{Ohmori:2014pca} and in an Appendix of \cite{Ohmori:2014kda}, respectively.
We can combine the two computations into one and one finds the following contribution to the anomaly, excluding the most subtle contribution from the codimension-5 singularity where the $\bC^2/\bZ_k$ singularity hits the end-of-the-world brane: \begin{multline}
I^\text{naive}_\text{inflow}(Q)= \frac{Q^3 k^2 }{6} c_2(R)^2 - \frac{ Q^2 k}{2} c_2(R) I_4 + \\
Q(\frac12 I_4^2-I_8) + (I_4-Q k c_2(R)) J_4 - \frac12 I^\text{vec}(\SU(k)) 
\end{multline}
where $Q$ is the M5-chage of the configuration,\begin{align}
I_8 &= \frac1{48}(p_2(N)+p_2(T)-\frac14(p_1(N)-p_1(T))^2),\\
I_4 &= \frac14(p_1(T) -2c_2(R)) ,\\
J_4 &= \frac1{48}(k-\frac1k)(4 c_2(R)+p_1(T)) + \frac14\tr F^2_{\SU(k)}.
\end{align}
Here $I_8$ comes from the M-theory interaction $\int C\wedge I_8$,
$I_4$ appears in the boundary condition $G=I_4$ at the $E_8$ wall,
and $J_4$ is the interaction on the $\bC^2/\bZ_k$ singular locus $\int C\wedge J_4$.
In this section the normalization of $\tr$ is as in \cite{Ohmori:2014pca}.

Let $\un{n}$ be the Kac label, and let $\vec w=\sum \vec w_i n_i$ be the corresponding weight vector.
By performing computations for many choices of $\un{n}$, we find that \begin{equation}
I_\text{quiver}(\un{n},\NN)=I^\text{naive}_\text{inflow}(Q)+c(\un{n})
\end{equation}
where $c(\un{n})$ is a constant depending on the Kac label $\un{n}$ but independent of $\NN$
and
\begin{equation}
Q=\NN+\frac12(k+\frac1k-\frac{\vev{\vec w,\vec w}}{k}).\label{M5charge}
\end{equation}
Recall that the instanton number as defined by the integral of $\tr F\wedge F$ was given by \begin{equation}
\int\tr F\wedge F =\Nprime -\frac{\vev{\vec w,\vec w}}{2k},
\end{equation} see \eqref{E8instantonnumber},
and that $k-1/k$ is the Euler number of $\widetilde{\bC^2/\Gamma}$, or equivalently of the integral of $-p_1/4$ there, see \eqref{chiGamma}.
Then, assuming that \begin{equation}
\Nprime =\NN+k,\label{NprimeN0}
\end{equation} we can rewrite the effective M5-brane charge $Q$ as
\begin{equation}
Q=\int\tr F\wedge F  + \int_{\widetilde{\bC^2/\bZ_k}} \frac{p_1}4
\end{equation}
which is what we expect from the curvature coupling on the E8 end-of-the-world brane.

The authors made a guess of the formula for $c(\un{n})$ by trial and error.
It has the form
\begin{equation}
	\begin{split}
		c(\un{n})=&\frac1k (P_0(\un{n})+P_2(\un{n})+P_4(\un{n})+P_6(\un{n}))+\frac12 I_\text{free vector}
	\end{split}
\end{equation}
where \begin{equation}
I_\text{free vector}=\frac1{5760}(-240 c_2(R)^2 - 120 c_2(R) p_1(T) - 7 p_1(T)^2 + 4 p_2(T))
\end{equation}
is the anomaly polynomial of a free vector multiplet and
$P_i(\un{n})$ is a homogeneous polynomial of $n_i$'s of degree $i$.
Those polynomials are identified as
\begin{align}
	P_0 =&\frac1{384} (-88 c_2(R)^2 + 32 c_2(R) p_1(T) - 5 p_1(T)^2 + 4 p_2(T))\\
	\begin{split}
		P_2 =& \frac1{11520} k^2\left(2512 c_2(R)^2 - 760 c_2(R) p_1(T) + 157 p_1(T)^2 - 124 p_2(T)\right)\\
	      &+\frac1{5760}\left(15 \langle \vec{w},\vec{w}\rangle-k\langle \vec{w},\vec{\rho}\rangle\right)\left(112 c_2(R)^2 - 40 c_2(R) p_1(T) + 7 p_1(T)^2 - 4 p_2(T)\right)
	\end{split}\\
	P_4 =&-\frac1{288}\left( 9 \langle \vec{w},\vec{w}\rangle^2 + 15 k^2 \langle \vec{w},\vec{w}\rangle - 2 k^4 -  k \sum_{\vec\alpha\in\Delta^+}\langle \vec{w},\vec\alpha\rangle^3\right)\left(4 c_2(R)^2 - c_2(R) p_1(T)\right) \\
	P_6 =&\frac1{240} \left(5 \langle  \vec{w},\vec{w}\rangle^3 + 15 k^2 \langle \vec{w},\vec{w}\rangle^2  - 5 k^4 \langle \vec{w},\vec{w}\rangle  + k^6 -  k\sum_{\vec\alpha\in\Delta^+}\langle \vec{w},\vec\alpha\rangle^5\right) c_2(R)^2,
\end{align}
where $\Delta^+$ is the set of positive roots of $E_8$.
The authors have not been able to determine how this formula come from the correct anomaly inflow calculation. It would be interesting to understand it.

\section{Lower dimensional incarnations}
\label{sec:lower}
\subsection{Five-dimensional brane-web description}
\label{sec:5d}
We can reduce the 6d theory on a circle to 5d. 
Roughly speaking, there are two different types of reductions.
For example, starting from the E-string theory, one can  obtain $\SU(2)$ theory with eight flavors in one way, or the 5d SCFT with $E_8$ flavor symmetry in the other way.

\paragraph{First reduction:}
Keeping the radius of the circle non-zero the low-energy 5d theory is sometimes a 5d gauge theory. 
Specifically, the class of 6d theories we are considering can be realized by a brane construction involving a system of NS5-branes and D6-branes in the presence of an O8$^-$-plane \cite{Brunner:1997gf,Hanany:1997gh}.
Performing T-duality on this system results in a brane configuration involving NS5-branes and D6-branes in the presence of an O8$^-$-plane. 
Alternatively, the system can also be described as D4-branes immersed in an O8$^-$-plane and D8-branes, in the presence of a $\bC^2/\bZ_k$ singularity \cite{BergmanGomez}.

Either way, the system can sometimes be deformed so as to describe a 5d gauge theory. 
Specifically, when compactifying we have a choice of the value of the radius as well as the freedom to turn on holonomies for the global symmetries. 
These then become mass parameters in the 5d theory. In specific ranges of these parameters the 6d theory may flow at low-energy to a 5d quiver gauge theory with the coupling constants of the gauge theory identified with the mass deformations. 
In general, a given 6d SCFT may have several different low-energy 5d gauge theory descriptions depending on the specific deformations used. 
Various 5d descriptions of 6d theories, including the type we are interested in, were studied in \cite{Zafrir:2015rga,Hayashi:2015zka,HayashiKimLee,HayashiKimLee2}. 
We will not consider this problem here.

\paragraph{Second reduction:}
Instead we shall take the limit of zero radius. In this case we argue that the 6d theory flows in the IR to a 5d SCFT. 
Furthermore, we claim that the 5d SCFT can be readily described in terms of the integer $N$ and the Kac label $\un{n}$. 
To find the 5d theory, we first  write down the 6d quiver following the algorithm presented in the last section.
We realize this 6d quiver in type IIA using O8-planes, D8-branes, D6-branes and NS5-branes as in \cite{Brunner:1997gf,Hanany:1997gh}.
We then compactify it on $S^1$, T-dualize it to type IIB, and manipulate the branes.
We will detail the procedure in slightly more detail below.

The result  can be conveniently represented by a brane web, which has a star shape form with a group of $(1,0)$, $(0,1)$ and $(1,1)$ 5-branes all intersecting at a point. 
The 5-branes end on the appropriate 7-branes where some collection of 5-branes end on the same 7-brane.
Specifying the configuration then is done by giving the distribution of 5-branes on the 7-branes. 
This is conveniently done by a Young diagram where each column represents a 7-brane, and the number of boxes in it represents the number of 5-branes ending on it. 

The three Young diagrams for the SCFTs we are considering are given by:
\begin{equation}
\begin{aligned}
Y_1=(&\nT-n_6,\\
&\nT-n_6-n_5,\\
&\nT-n_6-n_5-n_4,\\
&\nT-n_6-n_5-n_4-n_3,\\
&\nT-n_6-n_5-n_4-n_3-n_2,\\
&\nT-n_6-n_5-n_4-n_3-n_2-n_1,\\
&1^k),\\
Y_2=(&2\nT+2n_{4'}+n_{2'}+n_{3'},\\
&2\nT+n_{4'}+n_{2'}+n_{3'},\\
&2\nT+n_{4'}+n_{3'}),\\
Y_3=(&3\nT+2n_{4'}+n_{2'}+2n_{3'},\\
&3\nT+2n_{4'}+n_{2'}+n_{3'}).
\end{aligned}
\label{classSdata}
\end{equation}

\paragraph{More detail of the second reduction:}
For cases 1, 2 and 3, these results can be derived using the standard techniques.
But there are some issues for cases 4 and 5.
Case 4 naively does not have a brane construction of the type considered in \cite{Hanany:1997gh} so this procedure appears to be inapplicable in this case.
However, a conjecture for the 5d theories that lift to these types of 6d SCFTs was given in \cite{Zafrir:2015rga,Hayashi:2015zka}, and we can use this conjecture to fill in this step for case 4.

This leaves case 5.
We can ask how does the 6d $\theta$ angle appears in the brane construction.
In fact a similar issue arises in the analogue 5d system: D5-branes suspended between NS5-branes in the presence of an O7$^-$-plane.
In that case it was observed by \cite{BergmanZafrir} that accounting for the 5d $\theta$ angle seems to necessitate the introduction of two variants of the O7$^-$-plane, where one is an $SL(2,\bZ)$ T-transform of the other.
This in particular means that they differ by their decomposition into a pair of 7-branes.
Note that the distinction between the two cases vanishes when there are D7-branes on the O7$^-$-plane.
This becomes clear after we decompose the O7$^-$-plane into 7-branes which can be moved through the monodromy lines of the 7-branes which will change them by a T-transformation.
This of course agrees with the unphysical nature of the 5d $\theta$ angle once flavors are present.
There should be a similar distinction for the O8$^-$-plane, and so can account for the apparent 6d $\theta$ angle we observe.
We will not pursue this here.

However once we perform T-duality we end with a system with two O7$^-$-planes, and we expect that we can accommodate this in the observed difference in O7$^-$-planes.
We have a discrete choice for each O7$^-$-plane leading to four possibilities.
However we are free to perform a global T-transformation.
Since all the external branes are D7-branes, this will lead us to the same system, save for changing the types of both orientifolds.
Thus we conclude that there are only two distinct choices: the same or differing types.
These cases are expected to differ only when there are no 7-branes on the O7$^-$-planes, and thus no D8-brane on the original O8$^-$-plane.
This exactly agrees with the two cases, which coincide once $a_8 = 0$.
We indeed find different 5d theories for these two choices, where the former is identified with case 1 while the latter with case 5.
In this manner we can apply this procedure also to case 5. 

\subsection{Four-dimensional class S description}
\label{sec:4d}

We can compactify on an additional circle to 4d. 
Using the results of \cite{BBT}, it is straightforward to write the 4d theory. 
It is just an $A$ type class S theory given by the same set of Young diagrams as the 5d description,
given above in \eqref{classSdata}.

In fact it is also possible to motivate this class S description with the Young diagrams \eqref{classSdata} directly from the 4d description, and then use the preceding discussion to connect the 6d quiver data to the Kac labels. 
We start from the observation that the class S theory whose Young diagrams are \eqref{classSdata} can be thought of as generated by modifying the Young diagrams
of the rank $N$ $E_8$ theory, which is given by a class S theory of type $\SU(6N)$ with Young diagrams $Y_1=(N^6)$, $Y_2=(2N^3)$, $Y_3=(3N^2)$.

First the 4d theory needs to have the $\SU(k)$ global symmetry, coming from the $\bC^2/\bZ_k$ singularity. This is given by the $k$ boxes attached to the  Young diagram $Y_1$ of the $E_8$ theory. 
That this is the correct way to account for it can be seen by comparing anomalies. 
For the type of 6d theories we are considering, there is a result due to \cite{Ohmori:2015pua} that allows for the computations of the central charges of the 4d result of the compactification of the 6d theory from the anomaly polynomial of the latter. 
Furthermore the anomaly polynomial of the 6d theories of the type we considered was studied in \cite{Zafrir:2015rga}.
 When applied to our case we find that $k^\text{4d}_{\SU(k)} = 2k + 12$ independent of the details of the Kac label.
This agrees with the anomaly of the class S theory. 

In addition to the $\SU(k)$ we also have the commutant of the orbifold in $E_8$ as a global symmetry, which depends on the Kac labels. 
The $E_8$ global symmetry is accommodated by the Young diagram structure of the starting $E_8$ SCFT so it is natural to expect that modifying this will give the required global symmetry and take into account the Kac labels. 
The global symmetry which is manifest in the class S construction  is $\SU(2)\times \SU(3) \times \SU(6)$ which can be identified with the three legs of the affine Dynkin diagram. 
This becomes more apparent once we compactify to 3d and consider the mirror dual, which we consider more extensively in the next subsection. 

The point is that we can associate a node in the legs of the affine $E_8$ Dynkin diagram roughly with the difference between neighboring columns. 
The central node can be associated with the difference between the sum of the first columns of the three Young diagrams and the the total number of boxes in any of them. 
When that difference is zero, we get the $E_8$ theory. 
It is now natural to associate that difference to the Kac label of the corresponding node. 
By the Kac prescription, this ensures that we get the correct global symmetry. 
This leads to the conjectured form. 
There is one ambiguity in determining the total number of boxes which is related to the rank of the initial $E_8$ theory. 
This should be related to the number of tensors in 6d, but we need to determine the exact mapping. 
For this we use the relation outlined in the previous sections between the 6d and 4d theories.      

We can perform various consistency checks of this proposal. 
One check is to compare anomalies. 
We already mentioned that these can be computed from the 6d anomaly polynomial, and compare the $\SU(k)$ central charge. 
We can also compare the central charges $a$ and $c$, and the dimension of the Coulomb branch. 
These can then be calculated from the 6d quiver on one side, and from the class S theory on the other, in terms of the Kac labels and $\nT$. 
For the computations on the class S side, we use the standard results of \cite{Gaiotto:2009we,Chacaltana:2010ks} and reviewed e.g.~in \cite{Tachikawa:2015bga}.
The results themselves are rather complicated and not very illuminating, but we do find that all three objects agree between the two calculations.
Any interested reader can play around with the Mathematica file which comes with this paper to confirm this point.

\subsection{Three-dimensional star-shaped quiver description}
\label{sec:3d}
Let us now move on to the three dimensions. 
We translate the Young diagrams $Y_{1,2,3}$  given in \eqref{classSdata} which specify the class S punctures to the 3d mirror description using the results of \cite{Benini:2010uu}.
We find that the resulting theory is given by the quiver gauge theory \begin{equation}
\hat X:=
{\Node{}{1}}-{\Node{}{2}} -\cdots-\Node{}{k} -\node{}{\tilde N_1}-\node{}{\tilde N_2}-\node{}{\tilde N_3}-\node{}{\tilde N_4}-\node{}{\tilde N_5}-\node{\cver{}{\tilde N_{3'}}}{\tilde N_6}-\node{}{\tilde N_{4'}}-\node{}{\tilde N_{2'}}~.
\end{equation}
Here, all nodes are unitary with the diagonal $\UU(1)$  removed, 
and the gray and the black blobs are used as a visual aid for the affine Dynkin part and the over-extended part.
The ranks of the groups are specified by the vector
\begin{equation}
\un{\tilde N}=\NN \un{d}+\sum n_i \un{q_i} \label{N3d}
\end{equation}
where 
\begin{align}
\un{q_1}&=\E123456423, &
\un{q_2}&=\E223456423 ,&
\un{q_3}&=\E333456423, \\
\un{q_4}&=\E444456423 ,&
\un{q_5}&=\E555556423 , &
\un{q_6}&=\E666666423, \\
\un{q_{4'}}&=\E444444212, &
\un{q_{2'}}&=\E222222101,& 
\un{q_{3'}}&=\E333333211 
\end{align} which is in fact given by a uniform formula \begin{equation}
(q_{i})_j = d_i d_j - \vev{\vec w_i,\vec w_j}
\end{equation} where $\vec w_i$ is the weight vector for the node $i\neq 1$ and $\vec w_1=0$.

Another characterization of $\un{\tilde N}$ is \begin{equation}
C \un{\tilde N}= \E{k}00000000 + \un{n} \label{kn}
\end{equation} where $C$ is the affine Cartan matrix of $E_8$; this determines $\un{\tilde N}$ mod $\un{d}$.

The dimension of the Coulomb branch $\hat\cM$ is then \begin{equation}
\dim_\bH \hat\cM=30(\NN + k) - \vev{\vec w,\vec\rho} +\frac{k(k+1)}2-1
\end{equation} where $\vec w=\sum n_i \vec w_i$ is the Kac label as a weight vector and $\vec\rho=\sum_i \vec w_i$ is the Weyl vector.

The Coulomb branch $\hat\cM$ of this system $\hat X$ is closely related to the instanton moduli space $\cM^\text{inst}$ on the ALE space $\widetilde{\bC^2/\bZ_k}$.
To explain the relation, let us first recall that the resolution and deformation parameters of the ALE space can be specified by a parameter \begin{equation}
\xi=(\xi_\bC,\xi_\bR)\in \fsu(k)\otimes(\bC\oplus \bR)
\end{equation} which takes values in the Cartan of $\fsu(k)$ tensored by $\bR^3$.
We now need an auxiliary hyperk\"ahler space $\cO_{\xi}$, 
which is the $\SU(k)_\bC$ orbit of $\xi_\bC$ in $\fsu(k)$
with the hyperk\"ahler metric specified by $\xi_\bR$.
Equivalently, $\cO_\xi$  is the Coulomb/Higgs branch of the $T[\SU(k)]$ theory
whose quiver realization is given by 
\begin{equation}
T[\SU(k)]= \Node{}{1}-\Node{}{2}-\cdots-\Node{}{k-1}-\sqnode{}{k}
\end{equation} where the rightmost square node is a flavor symmetry
and $\xi$ is the $\SU(2)_R$ triplet of mass parameters associated to it.

We can now state the relation between $\hat\cM$ and and $\cM^\text{inst}$ by slightly modifying  an argument given in \cite{Ohmori:2015pua}: \begin{equation}
\cM^\text{inst}=(\hat\cM \times \cO_{\xi} )\hkq \SU(k) \label{cM}.
\end{equation} 
This relation can be understood as follows. 
The resolution/deformation parameter $\xi$ of the ALE space 
can be identified with the scalar vacuum expectation values of the 7d super $\SU(k)$ Yang-Mills theory supported on the M-theory singularity $\bC^2/\bZ_k$.
The 6d SCFT on the M5-branes at the intersection of the $E_8$ wall and the $\bC^2/\bZ_k$ singularity couples to this 7d super Yang-Mills, 
via the standard coupling where the triplet moment map field of the 6d theory is identified with the limiting value of the triplet of scalars of the 7d bulk.
The resulting hyperk\"ahler manifold is then given by the hyperk\"ahler reduction as in \eqref{cM}.

Now, our system $\hat X$ can also be written using the theory $\tilde X$ \begin{equation}
\tilde X:=\sqnode{}{k} -\node{}{\tilde N_1}-\node{}{\tilde N_2}-\node{}{\tilde N_3}-\node{}{\tilde N_4}-\node{}{\tilde N_5}-\node{\cver{}{\tilde N_{3'}}}{\tilde N_6}-\node{}{\tilde N_{4'}}-\node{}{\tilde N_{2'}}~.
\end{equation}  Indeed, \begin{equation}
\hat X= (T[\SU(k)] \times \tilde X)\hkq \SU(k)
\end{equation} 
where the symbol $T \hkq G$ means that we gauge the flavor symmetry $G$ of the theory $T$.

So the theory $X$ whose Coulomb branch is $\cM^\text{inst}$ in  \eqref{cM} is given by\begin{equation}
X=  (T[\SU(k)] \times T[\SU(k)] \times \tilde X)\hkq (\SU(k) \times \SU(k))
\end{equation}  But two $T[\SU(k)]$ gauged by a diagonal $\SU(k)$ is known to disappear, since it is the domain wall of 4d \Nequals4 SYM implementing the S-duality \cite{Gaiotto:2008ak}. So we have, in fact,\begin{equation}
X=\tilde X
\end{equation} and the ALE deformation parameter $\xi$ is now the mass parameter of the $\SU(k)$ flavor symmetry.
We have \begin{equation}
\dim_\bH \cM^\text{inst}= 30(\NN+k)-\vev{\vec w,\vec \rho}.\label{field-result}
\end{equation}
This nicely agrees with the computation from the geometry \eqref{geometry-result} by the identification \begin{equation}
\Nprime =\NN+k.\label{NprimeN}
\end{equation}
This relation between $\NN$ and $\Nprime $ is also consistent with what we found from the inflow, see \eqref{NprimeN0}.

We note that the theory $X=\tilde X$ is the theory whose Higgs branch is the $\UU(k)$ instanton moduli on $\bC^2/\Gamma_{E_8}$ \cite{KronheimerNakajima,Douglas:1996sw}. 
From this reason, the Coulomb branch, at least when the mass parameter is zero, has been conjectured to be the $E_8$ instanton moduli space on the singular space $\bC^2/\bZ_k$ by various people.
This follows, at least in a rough form, from the string duality: 
consider the theory on M2-branes on  $\bC^2/\bZ_k \times \bC^2/\Gamma_{E_8}$.
It has two supersymmetric branches of vacua, one describing $E_8$ instantons on $\bC^2/\bZ_k$ and another describing $\UU(k)$ instantons on $\bC^2/\Gamma_{E_8}$.
If the former is the Coulomb branch, then the latter is the Higgs branch.

Note that we arrived at the quiver gauge theory $X=\tilde X$ from a totally different method, by first studying the 6d quiver and then by reducing on successively on circles.
Therefore, this agreement can be thought of as an overall consistency check of our construction.

Now, applying  \cite{KronheimerNakajima} and \cite{Douglas:1996sw} in our case, we see that  the $\UU(k)$ holonomy at infinity of $\bC^2/\Gamma_{E_8}$ is  trivial, and the first Chern class $c_1$ satisfies $\int_{E_i} c_1 = n_i$ which can be read off from \eqref{kn}. 
It would be interesting to understand from M-theory point of view why the first Chern class  on the $\bC^2/\Gamma_{E_8}$ side is given by the asymptotic $E_8$ holonomy on the $\bC^2/\bZ_k$.
It seems important for the full story to consider a more general case where $\bC^2/\bZ_k$ is replaced by the multicenter Taub-NUT space, see e.g.~\cite{Bullimore:2015lsa,Nakajima:2015gxa}.

\section{Examples}
\label{sec:examples}
Let us demonstrate the above general statement in various examples.  
In this section, we take $N$ to be the number of tensor multiplets in the $6d$ theory, which was denoted by $\kappa$ in the other sections.

\subsection{The case of $k=2$}    
There are three possibilities. We label the cases with the Kac label $\un{n}$ and the group $H\subset E_8$ left unbroken by the Kac label.
The choice $k=2$ is somewhat special, since the ALE space $\widetilde{\bC^2/\bZ_2}$, also known as the Eguchi-Hanson space, has an exceptional isometry $\SU(2)$.
Then the generic flavor symmetry of the 6d SCFT should be $H\times \SU(2)^2$, where one $\SU(2)$ comes from the 7d gauge field on the singularity and another $\SU(2)$ comes from the isometry.
\ben
\item The first case is \begin{equation}
\un{n}=\E200000000, \qquad H=E_8.
\end{equation}
The corresponding $6d$ theory is
\be \label{6dE8inst}
[E_8] \, \, \, \, 1  \, \, \, \, \overset{\mathfrak{su}(1)}{2} \, \,\, \, \underset{\left[N_f=1 \right]}{\overset{\mathfrak{su}(2)}{2}} \, \,\, \, 
\underbrace{\overset{\mathfrak{su}(2)}{2} \quad \cdots\quad\overset{\mathfrak{su}(2)}{2}}_{N-3}\quad [\SU(2)].
\ee
The $T^3$ reduction of this theory gives the following $3d$ $\CN=4$ theory:
\be \label{compT2S16dE8inst}
{\Node{}{1}}-{\Node{}{2}} -\node{}{N}-\node{}{2N}-\node{}{3N}-\node{}{4N}-\node{}{5N}-\node{\cver{}{3N}}{6N}-\node{}{4N}-\node{}{2N}~.
\ee
\item The second case is  \begin{equation}
\un{n}=\E010000000, \qquad H=E_7\times \SU(2).
\end{equation}
The corresponding $6d$ theory is
\be 
[E_7] \,\,\,\, 1  \,\,\,\,  \underset{[N_f =2]}{\overset{\mathfrak{su}(2)}{2}} \,\,\,\,  \underbrace{\overset{\mathfrak{su}(2)}{2} \,\,\,\, \cdots  \,\,\,\,  \overset{\mathfrak{su}(2)}{2}}_{N-2} \,\,\,\, [\SU(2)]
\label{6dE7}
\ee
where the number of  $\mathfrak{su}(2)$ gauge groups in the quiver is $N-1$.  The Higgs branch dimension of the UV fixed point of this theory is $29 N +4 + 4(N-1) -3(N-1) = 30 N+3$.
The mirror of the $T^3$ compactification of this theory is
\be \label{quivcompB3E7}
\Node{}{1}-\Node{}{2}- \node{}{N+1}-\node{}{2N}-\node{}{3N}-\node{}{4N}-\node{}{5N}-\node{\cver{}{3N}}{6N}-\node{}{4N}-\node{}{2N}~,
\ee
The Coulomb branch dimension, which is the sum of the rank of the gauge groups minus one, is indeed $30N+3$.
\item The third case is \begin{equation}
\un{n}=\E000000010, \qquad H=\SO(16).
\end{equation}
The corresponding $6d$ theory is
\be \label{SO16k2}
[\SO(16)] \,\,\,\, \overset{\mathfrak{sp}(1)}{1}  \,\,\,\,  \underbrace{\overset{\mathfrak{su}(2)}{2} \,\,\,\,  \overset{\mathfrak{su}(2)}{2} \,\,\,\, \cdots  \,\,\,\,  \overset{\mathfrak{su}(2)}{2}}_{N-1} \,\,\,\, [\SU(2)]
\ee
where the number of $\SU(2)$ gauge groups associated with the $(-2)$ curves is $N-1$.
The Higgs branch dimension of the UV fixed point of this theory is $29 N + 16 + 4 + 4 (N - 1) - 3 - 3 (N - 1) = 30N+16$.
The mirror of the $T^3$ compactification of this theory is
\be \label{3dquivSO16k2}
\Node{}{1}-\Node{}{2}-\node{}{N+2}-\node{}{2N+2}-\node{}{3N+2}-\node{}{4N+2}-\node{}{5N+2}-\node{\cver{}{3N+1}}{6N+2}-\node{}{4N+1}-\node{}{2N}.
\ee
The Coulomb branch dimension, which is the sum of the rank of the gauge groups minus one, is indeed $30N+16$.
This is consistent with Figure 45 of \cite{Zafrir:2015rga}, namely the $T^2$ compactification of \eref{SO16k2} yields the class $\CS$ theory whose Gaiotto curve is a sphere with punctures:
\be
[(3N+1)^2], \qquad [(2N+1)^2,2N], \qquad [N^6,1^2]~.
\ee
\een

Now let us comment on the flavor symmetry from the point of view of the 6d quiver.
Since an $\SU(2)$-$\SU(2)$ bifundamental has an $\SU(2)$ flavor symmetry,
the three 6d quivers presented above have order $N$ copies of $\SU(2)$ symmetries on the generic points of the tensor branch.
In fact the same issue already appears in the case of $N$ M5-branes probing the $\bC^2/\bZ_2$ singularity, which has the quiver \begin{equation}
[\SU(2)] \quad
\underbrace{
\overset{\fsu(2)}{2}\quad
\cdots\quad
\overset{\fsu(2)}{2}
}_{N}\quad
[\SU(2)]
\end{equation} which naively has too many $\SU(2)$ flavor symmetries.

The issue can be resolved by recalling the fact derived in Appendix A of \cite{Ohmori:2015pia} that  the basic 6d SCFT whose quiver on the tensor branch is given by $\SU(2)$ with $N_f=4$ with a naive $\SO(8)$ symmetry, only has an $\SO(7)$ symmetry under which the flavors transform in the spin representation. 
In the quiver representation of the same theory as \begin{equation}
[\SU(2)_1]\quad \overset{\fsu(2)}{2} \quad [\SU(2)_2],\label{Z2}
\end{equation}
this means the following: 
regard the bifundamental hypermultiplets on the left and on the right of the gauge group as the trifundamental half-hypermultiplets.
At the quiver level there are therefore the flavor symmetry $\SU(2)_1\times \SU(2)_1'\times \SU(2)_2\times \SU(2)_2' \subset \SO(8)$.
Under the $\SO(7)$ symmetry which is the flavor symmetry of the SCFT, only the diagonal subgroup of $\SU(2)_1'$ and $\SU(2)_2'$ survives.
Applying this argument at every $\fsu(2)$ node in \eqref{6dE8inst}, \eqref{6dE7}, \eqref{SO16k2}, and \eqref{Z2}, we see that the number of $\SU(2)$ flavor symmetries is reduced appropriately.

There are also some interesting special cases with enhanced flavour symmetries when $N$ is small:
\ben
\item $N=2$, $H=E_8$. 
In this case the quiver \eqref{6dE8inst} degenerates to 
\begin{equation}
[E_8] \, \, \, \, 1  \, \, \, \, \overset{\mathfrak{su}(1)}{2} \, \,\, \, \underset{\left[N_f=1 \right]}
{[\SU(2)]}
\end{equation}
which is just the rank-2 E-string theory with three decoupled hypermultiplets.
The 3d quiver in this case is \eqref{compT2S16dE8inst} for $N=2$:
\begin{equation}
{\Node{}{1}}-{\Node{}{2}} -\node{}{2}-\node{}{4}-\node{}{6}-\node{}{8}-\node{}{10}-\node{\cver{}{6}}{12}-\node{}{8}-\node{}{4}~.
\end{equation}
Its Coulomb branch is 
\be \label{CoulHE8}
\bH^3 \times (\text{the reduced moduli space of 2 $E_8$ instantons on $\BC^2$})
\ee
and we indeed see the same decoupled structure.  The explanation from the perspective of the Coulomb branch operators will be described below.

\item $N=3$ with $H=E_8$. 
The 6d quiver is \begin{equation}
[E_8] \, \, \, \, 1  \, \, \, \, \overset{\mathfrak{su}(1)}{2} \, \,\, \, \underset{\left[N_f=1 \right]}{\overset{\mathfrak{su}(2)}{2}}  \, \,\, \, [\SU(2)].
\end{equation}
The 3d quiver is 
\be
{\Node{}{1}}-{\Node{}{2}}-\node{}{3}-\node{}{6}-\node{}{9}-\node{}{12}-\node{}{15}-\node{\cver{}{9}}{18}-\node{}{12}-\node{}{6}
\ee
The flavour symmetry is enhanced to $G_2 \times E_8$.
The explanation from the perspective of the Coulomb branch operators will also be described below.

\item $N=2$, $H= E_7 \times \SU(2)$.  
The 6d quiver for this case reduces to 
\begin{equation} \label{E7SU2N2k2}
[E_7] \,\,\,\, 1  \,\,\,\,  \underset{[N_f=2]}{\overset{\mathfrak{su}(2)}{2}}   \,\,\,\, [\SU(2)].
\end{equation}
On the tensor branch, there is an $\SO(8)$ symmetry acting on the four flavors of $\SU(2)$ gauge group.
In the SCFT it is known that there is only $\SO(7)$.
The total symmetry is then $\SO(7)\times E_7$.
In fact this 6d theory is the $(E_7, \SO(7))$ minimal conformal matter~\cite{DelZotto:2014hpa}, which describes ``half M5-branes'' on the $E_7$ singularity. 

The 3d quiver in this case is \eqref{quivcompB3E7} for $N=2$:
\begin{equation}
\Node{}{1}-\Node{}{2}- \node{}{3}-\node{}{4}-\node{}{6}-\node{}{8}-\node{}{10}-\node{\cver{}{6}}{12}-\node{}{8}-\node{}{4}~.
\end{equation}
This theory is the mirror of the $S^1$ reduction of the class $\CS$ theory whose Gaiotto curve is a sphere with punctures
\be \label{classShalfM5onE7}
[2^4, 1^4], \qquad [6^2], \qquad [4^3]~.
\ee 

In  \cite{Zafrir:2015rga,Ohmori:2015pua} the $T^2$ compactification was also identified with  a class $\CS$ theory of the $E_6$ type associated with the sphere with punctures $0$, $2A_1$ and $E_6(a_1)$.  For consistency, these two class S theories should in fact be the same.
Let us compute the central charges of \eref{classShalfM5onE7}.  We find that the effective numbers of vector multiplets and hypermultiplets are $n_H= 112$ and $n_V = 49$, respectively.  Thus,
\be
a = \frac{1}{24}( 5 n_V+n_H) = \frac{119}{8}~, \qquad c = \frac{1}{12}(2n_V + n_H) = \frac{35}{2}~.
\ee
This agrees with $a$ and $c$ of the aforementioned class $\CS$ theory of the $E_6$ type; see (7.1) of \cite{Ohmori:2015pua}.

\een

\subsection{Enhanced flavor symmetries from 3d quivers}  

In fact the symmetry enhancement of each of the three cases above can be generalized to other over-extended Dynkin quivers in 3d, namely:
\begin{enumerate}
\item For the quiver consisting of a tail ${\Node{}{1}}-{\Node{}{2}}$ attached to the affine Dynkin diagram of type $\mathfrak{g}$ with gauge groups being unitary groups of the ranks given by 2 times the dual Coxeter labels, the Coulomb branch moduli space is $\bH^3 \times \widetilde{\CM}_{2, \mathfrak{g}}$, where $\widetilde{\CM}_{2, \mathfrak{g}}$ denotes the reduced two-instanton moduli space of group $\mathfrak{g}$  on $\BC^2$.  For example, the Coulomb branch of the quiver 
\be 
{\Node{}{1}}-{\Node{}{2}} -{\node{}{2}}= {\node{}{2}}
\ee 
is $\bH^3 \times \widetilde{\CM}_{2,\,\, \mathfrak{su}(2)}$, and  the Coulomb branch of the quiver 
\be
{\Node{}{1}}-{\Node{}{2}} -{\node{}{2}}-\underset{\uer{}{2}}{\node{\cver{}{2}}{4}}-\node{}{2}
\ee
is $\bH^3 \times \widetilde{\CM}_{2, \,\, \mathfrak{so}(8)}$.   

\item For the quiver consisting of a tail ${\Node{}{1}}-{\Node{}{2}}$ attached to the affine Dynkin diagram of type $\mathfrak{g}$ with gauge groups being unitary groups of the ranks given by 3 times the dual Coxeter labels, the Coulomb branch moduli space has a symmetry $G_2 \times \mathfrak{g}$.  For example, the Coulomb branch of the quiver 
\be {\Node{}{1}}-{\Node{}{2}} -{\node{}{3}}= {\node{}{3}}
\ee 
has a symmetry $G_2 \times \SU(2)$, and the Coulomb branch of the quiver 
\be {\Node{}{1}}-{\Node{}{2}} -{\node{}{3}}-\underset{\uer{}{3}}{\node{\cver{}{3}}{6}}-\node{}{3}
\ee
has a symmetry $G_2 \times \SO(8)$.  

\item For the quiver consisting of a tail ${\Node{}{1}}-{\Node{}{2}}$ attached to the affine Dynkin diagram of type $\mathfrak{g}$ with the affine node being $U(3)$ and other gauge groups being unitary groups of the ranks given by 2 times the dual Coxeter labels, the Coulomb branch has  a symmetry $\SO(7) \times \tilde{\mathfrak{g}}$, where $\tilde{\mathfrak{g}}$ is the commutant of $\mathfrak{su}(2)$ in $\mathfrak{g}$.  For example, the Coulomb branch of the following quiver
\be
{\Node{}{1}}-{\Node{}{2}}-{\node{}{3}}-{\node{}{4}}-\node{\overset{\cver{}{2}}{\cver{}{4}}}{6}-{\node{}{4}}-{\node{}{2}}
\ee
has a symmetry $\SO(7) \times \SU(6)$, where $\SU(6)$ is the commutant of $\SU(2)$ in $E_6$. 

\end{enumerate}

In each of the above examples, the quiver contains of a balanced affine Dynkin quiver diagram as a subquiver.  If we consider only this subquiver, the R-charges of the monopole operators in this theory vanish, and hence this subquiver is indeed a bad theory.  By attaching a quiver tail ${\Node{}{1}}-{\Node{}{2}}-\cdots-{\Node{}{k}}$ to such a subquiver, the total quiver becomes good or ugly.\footnote{See also \cite{Gaiotto:2012uq} for a related consideration from the $4d$ point of view.}  We would like to consider the contribution of this quiver tail to the Coulomb branch of the total quiver.  

\begin{enumerate}
\item For this case, the node $\node{}{2}$, which is the affine node in the affine Dynkin diagram, is over-balanced in the sense of \cite{Gaiotto:2008ak}.  Following \cite{Gaiotto:2008ak}, we can split the quiver into two parts, namely ${\Node{}{1}}-{\Node{}{2}} -\node{}{2}$ and the rest of the Dynkin diagram.  The R-charge of the monopole operators from the subquiver ${\Node{}{1}}-{\Node{}{2}} -\node{}{2}$ receives the contribution from the hypermultiplets and vector multiplets in the way described in \cite{Gaiotto:2008ak}, except that there is no contribution from the vector multiplet of the rightmost node $\node{}{2}$, since this was cancelled inside the affine Dynkin quiver.  The contribution from the subquiver is therefore the same as that of the quiver ${\Node{}{1}}-{\Node{}{2}} -\node{\bigcap}{2}$, where $\cap$ denotes an adjoint hypermultiplet of the $\UU(2)$ rightmost node.  The Coulomb branch of ${\Node{}{1}}-{\Node{}{2}} -\node{\bigcap}{2}$ contains 3 free hypermultiplets, which can be seen from the monopole operators with $\SU(2)_R$-spin $1/2$.  This explains the $\BH^3$ factor in \eref{CoulHE8}.  The reduced moduli space of two $E_8$ instantons on $\BC^2$ can be realised as in \cite{Cremonesi:2014xha}.

\item Similarly, for this case,  the total quiver can be split into ${\Node{}{1}}-{\Node{}{2}} -\node{}{3}$ and the rest of the Dynkin diagram.  The contribution to the R-charge of the monopole operators from the subquiver ${\Node{}{1}}-{\Node{}{2}} -\node{}{3}$ can be realised from the quiver ${\Node{}{1}}-{\Node{}{2}} -\node{\bigcap}{3}$, where $\cap$ denotes an adjoint hypermultiplet of the $\UU(3)$ rightmost node\footnote{The authors thank S. Cremonesi for this argument.}.  Indeed, it was pointed out in section 3.3.2 of \cite{Cremonesi:2014vla} that the Coulomb branch of the latter model has a $G_2$ symmetry.  (Note that the corresponding $4d$ class S theory had been studied in \cite{Gaiotto:2012uq}. The $G_2$ symmetry on the Higgs branch of such a theory had also been pointed out in that reference.) This therefore explains the $G_2$ symmetry in case 3. The $E_8$ symmetry follows from the Dynkin subquiver.

\item Finally, for this case, \,\, $\node{}{4}$\,\, is the unbalanced node in the quiver.  There are two contributions to the Coulomb branch operators with $\SU(2)_R$-spin $1$.  One contribution can be realised using the quiver ${\Node{}{1}}-{\Node{}{2}}-{\node{}{3}} -\node{\bigcap}{4}$ in a similar fashion to the above discussion.  This quiver has a Coulomb branch symmetry $\SU(4)$ and thus gives $15$ operators with $\SU(2)_R$-spin $1$ in the adjoint representation of $\SU(4)$.   The other contribution can be seen as follows.  Since the node ${\node{}{3}}$, which was originally a part of the affine Dynkin subquiver, now belongs to the tail ${\Node{}{1}}-{\Node{}{2}}-{\node{}{3}} -\node{}{4}$, we also need to take into account the contribution that arises from the removal of this node from such an affine Dynkin diagram.  The second contribution thus comes from considering ${\Node{}{1}}-{\Node{}{2}}-\node{\bigcap}{4}$.  There are $6$ Coulomb branch operators with $\SU(2)_R$-spin $1$ in the latter. Therefore, we have in total $15+6=21$ operators with $\SU(2)_R$-spin $1$; this explains the enhancement to the $\SO(7)$ symmetry.  The remaining symmetry is thus the commutant of $\SU(2)$, which arises from node ${\node{}{3}}$, in the original symmetry associated with the affine Dynkin diagram.

\end{enumerate}

\subsection{The case of $k=4$}  
There are ten possibilities.  The F-theory quiver for the $6d$ theories are listed on Page 73 of \cite{Heckman:2015bfa}.  Here are the mirrors of the $T^3$ compactification of them.   
\ben
\item The first case is \begin{equation}
\un{n}=\E400000000, \qquad H=E_8.
\end{equation} 
The 6d quiver is
\be
[E_8]\,\, 1  \,\, \overset{\mathfrak{su}(1)}2 \,\, \overset{\mathfrak{su}(2)}2 \,\,\overset{\mathfrak{su}(3)}2 \,\, 
\overbrace{\underset{[N_f =1]}{\overset{\mathfrak{su}(4)}2} \,\, \cdots\,\,\overset{\mathfrak{su}(4)}2}^{N-4} \,\, [\SU(4)]
\ee
and the 3d quiver is
\be
{\Node{}{1}}-{\Node{}{2}}-{\Node{}{3}} -{\Node{}{4}}-\node{}{N}-\node{}{2N}-\node{}{3N}-\node{}{4N}-\node{}{5N}-\node{\cver{}{3N}}{6N}-\node{}{4N}-\node{}{2N}.
\ee
\item 
The second case is \begin{equation}
\un{n}=\E210000000, \qquad H=E_7\times \UU(1)
\end{equation} 
with the 6d quiver
\be
[E_7]\,\, 1 \,\,\underset{[N_f=1]}{\overset{\mathfrak{su}(2)}2} \,\,\overset{\mathfrak{su}(3)}2 \,\, \overbrace{\underset{[N_f=1]}{\overset{\mathfrak{su}(4)}2} \,\, ... \,\, \overset{\mathfrak{su}(4)}2}^{N-3} \,\, [\SU(4)].
\ee
The dimension of the SCFT Higgs branch is 
\be
29N + 2 + 6 + 12 + 4 + 16 (N - 3) - 3 - 8 - 15 (N - 3) = 30N+10~.
\ee
The 3d quiver is
\be
{\Node{}{1}}-{\Node{}{2}}-{\Node{}{3}} -{\Node{}{4}}-\node{}{N+1}-\node{}{2N}-\node{}{3N}-\node{}{4N}-\node{}{5N}-\node{\cver{}{3N}}{6N}-\node{}{4N}-\node{}{2N}
\ee
and the dimension of the Coulomb branch is $30N+10$.
\item 
The third case is \begin{equation}
\un{n}=\E200000010, \qquad H=\SO(14)\times \UU(1)
\end{equation} 
with the 6d quiver
\be
[\SO(14)]\,\, \overset{\mathfrak{sp}(1)}1 \,\,\overset{\mathfrak{su}(3)}2 \,\,\overbrace{\underset{[N_f=1]}{\overset{\mathfrak{su}(4)}2} \,\, ... \overset{\mathfrak{su}(4)}2}^{N-2} \,\, [\SU(4)].
\ee
The dimension of the SCFT Higgs branch is
\be
29N + 14 + 6 + 12 + 4 + 16 (N - 2) - 3 - 8 - 15 (N - 2) = 30N+23~.
\ee
The 3d quiver is
\be
{\Node{}{1}}-{\Node{}{2}}-{\Node{}{3}} -{\Node{}{4}}-\node{}{N+2}-\node{}{2N+2}-\node{}{3N+2}-\node{}{4N+2}-\node{}{5N+2}-\node{\cver{}{3N+1}}{6N+2}-\node{}{4N+1}-\node{}{2N}
\ee
and the Coulomb branch dimension is $30N+23$.
\item The fourth case is \begin{equation}
\un{n}=\E020000000, \qquad H=E_7\times \SU(2).
\end{equation} 
with the 6d quiver
\be
[E_7]\,\, 1 \,\,\overset{\mathfrak{su}(2)}{2} \,\,\overbrace{\underset{[\SU(2)]}{\overset{\mathfrak{su}(4)}2} \,\, ... \overset{\mathfrak{su}(4)}2}^{N-2} \,\, [\SU(4)].
\ee
The dimension of the SCFT Higgs branch is
\be
29 N+ 8 + 8 + 16 (N - 2) - 3 - 15 (N - 2) = 30N+11~.
\ee
The 3d mirror is
\be
{\Node{}{1}}-{\Node{}{2}}-{\Node{}{3}} -{\Node{}{4}}-\node{}{N+2}-\node{}{2N}-\node{}{3N}-\node{}{4N}-\node{}{5N}-\node{\cver{}{3N}}{6N}-\node{}{4N}-\node{}{2N}
\ee
The Coulomb branch dimension is $30N+11$.
\item The fifth case is \begin{equation}
\un{n}=\E000000020, \qquad H=\SO(16)
\end{equation} with the 6d quiver
\be
[\SO(16)]\,\, \overset{\mathfrak{sp}(2)}1 \,\,\overbrace{\overset{\mathfrak{su}(4)}2  \,\, ... \overset{\mathfrak{su}(4)}2}^{N-1} \,\, [\SU(4)]
\ee
The dimension of the SCFT Higgs branch is
\be
29 N + 32 + 16 + 16 (N - 1) - 10 - 15 (N - 1) = 30N +37~.
\ee
The 3d quiver is
\be
{\Node{}{1}}-{\Node{}{2}}-{\Node{}{3}} -{\Node{}{4}}-\node{}{N+4}-\node{}{2N+4}-\node{}{3N+4}-\node{}{4N+4}-\node{}{5N+4}-\node{\cver{}{3N+2}}{6N+4}-\node{}{4N+2}-\node{}{2N}
\ee
and the Coulomb branch dimension is $30N+37$.
\item The sixth case is\begin{equation}
\un{n}=\E010000010, \qquad H=\SO(12)\times\SU(2)\times \UU(1)
\end{equation} 
with the 6d quiver
\be
[\SO(12)]\,\, \overset{\mathfrak{sp}(1)}1 \,\,\overbrace{\overset{\mathfrak{su}(4)}{\underset{[\SU(2)]}2}  \,\, ... \overset{\mathfrak{su}(4)}2}^{N-1} \,\, [\SU(4)]
\ee
The dimension of the SCFT Higgs branch is
\be
29 N + 12 + 8 + 8 + 16 (N - 1) - 3 - 15 (N - 1) = 30N +24~.
\ee
The 3d quiver is
\be
{\Node{}{1}}-{\Node{}{2}}-{\Node{}{3}} -{\Node{}{4}}-\node{}{N+3}-\node{}{2N+2}-\node{}{3N+2}-\node{}{4N+2}-\node{}{5N+2}-\node{\cver{}{3N+1}}{6N+2}-\node{}{4N+1}-\node{}{2N}
\ee
The Coulomb branch dimension is $30N+24$.
\item The seventh case is \begin{equation}
\un{n}=\E101000000, \qquad H=E_6 \times \SU(2) \times \UU(1).
\end{equation} 
with the 6d quiver
\be
[E_6]\,\, 1 \,\,\overset{\mathfrak{su}(3)}{\underset{[\SU(2)]}2} \,\,\overbrace{\underset{[N_f=1]}{\overset{\mathfrak{su}(4)}2} \,\, ... \overset{\mathfrak{su}(4)}2}^{N-2} \,\, [\SU(4)].
\ee
The dimension of the SCFT Higgs branch is
\be
29 N + 6 + 12 + 4 + 16 (N - 2) - 8 - 15 (N - 2) = 30N +12~.
\ee
The 3d mirror is
\be \label{30Nplus12}
{\Node{}{1}}-{\Node{}{2}}-{\Node{}{3}} -{\Node{}{4}}-\node{}{N+2}-\node{}{2N+1}-\node{}{3N}-\node{}{4N}-\node{}{5N}-\node{\cver{}{3N}}{6N}-\node{}{4N}-\node{}{2N}
\ee
and the Coulomb branch dimension is $30N+12$.
\item The eighth case is \begin{equation}
\un{n}=\E100000001,\qquad H=\SU(8)\times\UU(1)
\end{equation}
with the 6d quiver
\be
[\SU(8)]\,\, \overset{\mathfrak{su}(3)}1 \,\, \overbrace{\underset{[N_f=1]}{\overset{\mathfrak{su}(4)}2}  \,\, ... \overset{\mathfrak{su}(4)}2}^{N-1} \,\, [\SU(4)].
\ee
The dimension of the SCFT Higgs branch is
\be
29 N + 24 + 12 + 4 + 16 (N - 1) - 8 - 15 (N - 1) = 30N +31~.
\ee
The 3d quiver is
\be
{\Node{}{1}}-{\Node{}{2}}-{\Node{}{3}} -{\Node{}{4}}-\node{}{N+3}-\node{}{2N+3}-\node{}{3N+3}-\node{}{4N+3}-\node{}{5N+3}-\node{\cver{}{3N+1}}{6N+3}-\node{}{4N+2}-\node{}{2N+1}
\ee
and the Coulomb branch dimension is $30N+31$.
\item The ninth case is \begin{equation}
\un{n}=\E000100000,\qquad H=\SO(10)\times\SU(4)
\end{equation}
with the 6d quiver
\be
[\SO(10)]\,\, 1 \,\,\overbrace{\overset{\mathfrak{su}(4)}{\underset{[\SU(4)]}2}  \,\, ... \overset{\mathfrak{su}(4)}2}^{N-1} \,\, [\SU(4)].
\ee
The dimension of the SCFT Higgs branch is
\be
29 N + 16 + 16 (N - 1) - 15 (N - 1) = 30+15~.
\ee
The 3d quiver is
\be
{\Node{}{1}}-{\Node{}{2}}-{\Node{}{3}} -{\Node{}{4}}-\node{}{N+3}-\node{}{2N+2}-\node{}{3N+1}-\node{}{4N}-\node{}{5N}-\node{\cver{}{3N}}{6N}-\node{}{4N}-\node{}{2N}
\ee
and the dimension of the Coulomb branch is $30N+15$.
\item The final tenth case is \begin{equation}
\un{n}=\E000000100,\qquad H=\SU(8)\times\SU(2),
\end{equation}
with the 6d quiver
\be
[\SU(8)]\,\,  \overset{\mathfrak{su}(4)}{\underset{\text{[antisym]}}1} \,\,\overbrace{\overset{\mathfrak{su}(4)}2  \,\, ... \overset{\mathfrak{su}(4)}2}^{N-1} \,\, [\SU(4)]
\ee
The dimension of the SCFT Higgs branch is
\be
29 N + 32 + 6 + 16 N - 15 N = 30N+38~.
\ee
The 3d quiver is
\be
{\Node{}{1}}-{\Node{}{2}}-{\Node{}{3}} -{\Node{}{4}}-\node{}{N+4}-\node{}{2N+4}-\node{}{3N+4}-\node{}{4N+4}-\node{}{5N+4}-\node{\cver{}{3N+2}}{6N+4}-\node{}{4N+2}-\node{}{2N+1}
\ee
and the dimension of the Coulomb branch is $30N+38$.
\een

\subsection{Theories differing by the $6d$ $\theta$ angle}

In this subsection we look at the $4d$ and $3d$ theories generated from $6d$ SCFTs differing by the choice of $6d$ $\theta$ angle. 
The first case where this possibility occurs is for $k=8$, where the two choices are given by Kac labels $n'_3=2, n'_2=1$ for one and $n'_4=2$ for the other with the rest zero.
These can be generalized to $k=2l+8$ with Kac labels $n'_3=2, n'_2=1+l$ for one and $n'_4=2, n'_2=l$ for the other with the rest zero.
The $6d$ quiver in both cases is given by:

\be
\overset{\mathfrak{usp}(2l)}1 \,\,\overbrace{\overset{\mathfrak{su}(2l+8)}{\underset{[\SU(8)]}2}  \,\, ... \overset{\mathfrak{su}(2l+8)}2}^{N-1} \,\, [\SU(2l+8)]
\ee
where we identify the case $n'_4=2, n'_2=l$ with $\theta=0$ and $n'_3=2, n'_2=1+l$ with $\theta=\pi$. 

The associated $4d$ theories are different for the two cases.
In the $\theta=0$ case we associate the class S theory given by:

\be
[(N-1)^6, 1^{2l+8}], \quad [2N+l+2, 2N+l,2N], \quad [(3N+l+1)^2]~,
\ee
while the $\theta=\pi$ case is associated with:

\be
[(N-1)^6, 1^{2l+8}], \quad [(2N+l+1)^2,2N], \quad [3N+l+2,3N+l]~.
\ee

The $3d$ quivers are:

\be
{\Node{}{1}}-{\Node{}{2}}- ... -{\Node{}{2l+7}}-{\Node{}{2l+8}} -\node{}{N+2l+7}-\node{}{2N+2l+6}-\node{}{3N+2l+5}-\node{}{4N+2l+4}-\node{}{5N+2l+3}-\node{\cver{}{3N+l+1}}{6N+2l+2}-\node{}{4N+l}-\node{}{2N} ,
\ee
for the $\theta=0$ case, and

\be
{\Node{}{1}}-{\Node{}{2}}- ... -{\Node{}{2l+7}}-{\Node{}{2l+8}} -\node{}{N+2l+7}-\node{}{2N+2l+6}-\node{}{3N+2l+5}-\node{}{4N+2l+4}-\node{}{5N+2l+3}-\node{\cver{}{3N+l}}{6N+2l+2}-\node{}{4N+l+1}-\node{}{2N} ,
\ee
for the $\theta=\pi$ case.

We can now inquire as to how these theories differ from one another. 
In the $l=0$ case they differ already at the level of the global symmetry, where the $\theta=0$ case has an $\SU(8)^2 \times \SU(2) \times \UU(1)$ global symmetry while the $\theta=\pi$ case has an $\SU(8)^2 \times \UU(1)^2$ global symmetry.
In this case we have an $\SU(8)$ gauging of $E_8$ and the two choices differ by their commutant inside $E_8$.
We note that this difference is in accordance with the symmetry expected from the Kac labels.
When $l>0$ the symmetries of the two theories agree.

We can calculate the $4d$ anomalies of the two theories and find that all of them agree between the two theories.
Again this is consistent with our interpretation as the $4d$ anomalies can be computed from their $6d$ counterparts, which in turn are independent of the $\theta$ angle.
From our $6d$ interpretation we expect the two to differ slightly in their operator spectrum.
Particularly the $\theta$ angle should affect the USp gauge group instanton strings changing their charges under the global and gauge symmetries.
Upon compactification to lower dimensions these should map to local operators.

We can observe this from the $3d$ quivers.
We get a tower of monopole operators from every node.
The basic monopole operator from the balanced nodes leads to enhancement of symmetry.
We also have a basic monopole operator from the unbalanced nodes.
These provide operators with higher R-charges, and we can read of their R-charges and non-abelian global symmetry charges from the quiver.

We have three unbalanced nodes.
Two of them give the same contribution in both theories: one operator of $\SU(2)_R$ spin $\frac{N}{2}$ in the bifundamental of the $\SU(2l+8)\times \SU(8)$ global symmetry, and one operator of $\SU(2)_R$ spin $2$ in the $\bold{28}$ of the $\SU(8)$ global symmetry.
These can be readily identified with gauge invariants in the $6d$ quiver, where the former is the one made from $N-2$ $\SU(2l+8)\times \SU(2l+8)$ bifundamentals and the flavors, and the later is made from two $\SU(8)$ flavors and the $\USp(2l)\times \SU(2l+8)$ bifundamental.
The last one differ slightly between the two theories.

In the $\theta=0$ case it is a flavor singlet with $\SU(2)_R$ spin $\frac{l+2}{2}$.
Particularly for $l=0$ this gives the conserved current enhancing the $\UU(1)$ to $\SU(2)$.
In the $\theta=\pi$ case, however, it is in the $\bold{8}$ of $\SU(8)$ with $\SU(2)_R$ spin $\frac{l+3}{2}$.
We can interpret these states as coming from the $\USp$ gauge group instanton strings wrapped on the circle.
These are in the spinor of $\SO(4l+16)$, and depending on the $\theta$ angle decompose to all the even or odd rank antisymmetric tensor representations of the gauge $\SU(2l+8)$ connected to the $\USp$ gauge group.
In the $\theta=0$ case we get the even rank representations, which contain a gauge invariant part which is a flavor symmetry singlet.
In the $\theta=\pi$ case we get the odd rank representations, which do not contain any gauge invariants.
However we can combine it with one of the $\SU(2l+8)$ flavors to form an invariant.
This should contribute a state in the $\bold{8}$ of $\SU(8)$ with $\SU(2)_R$ spin which is greater by $\frac{1}{2}$ from that of the singlet.
This agrees with what we observe.
It might be interesting to study more accurately the spectrum, particularly, the Higgs branch chiral ring, and compare against the $6d$ expectations.
We will not pursue this here. 

\def\qappa{\varpi}
\subsection{Massive E-string theories}
\begin{table}
\begin{center}
\begin{tabular}{|c|c|c|c|}
\hline
$m_0$ & $E_{9-m_0}$ & Kac label $\un{n}/\qappa$ & $r_i/\qappa$ \\
\hline
\hline
$1$ & $E_8$ & $\E 100000000$ & $\E000000000$ \\
\hline
$2$ & $E_7 $ & $\E {0} {1} 0000000$ & $\E {1}00000000$ \\
\hline
$3$ & $E_6$ & $\E {0} {0} {1} 000000$ & $\E {2}{1}0000000$ \\
\hline
$4$ & $\SO(10)$ & $\E {0} {0} {0} {1} 00000$  & $\E{3} {2} {1}000000$ \\
\hline
$5$ & $\SU(5)$ & $\E {0} {0} {0} {0}{1} 0000$ & $\E{4} {3} {2} {1} 00000$\\
\hline
$6$ & $\SU(3) \times \SU(2)$  & $\E {0} {0} {0} {0}{0}{1} 000$ & $\E{5} {4} {3} {2} {1} 0000$ \\
\hline
$7$ & $\SU(2) \times \UU(1)$ & $\E {0} {0} {0} {0}{0}{0}  {1} 0 {1}$ & $\E {6} {5} {4} {3} {2} {1} {0} {0} {0}$ \\
\hline
$8$ & $\SU(2)$ & $\E {0} {0} {0} {0}{0}{0}  {2} 0 0$ & $\E {7} {6} {5} {4} {3} {2} {0} {0} {1}$ \\
\hline
\end{tabular}
\end{center}
\caption{The values of $r_i$ in \eref{mirgenmassEstring} and the Kac label for each $m_0$.}
\label{tab:m0ri}
\end{table}%

In this subsection, we consider the following $6d$ theory
\be \label{genmassEstring}
\CT^{6d}_E(\qappa ,m_0, N): \qquad [E_{9-m_0}] \,\,
	{1} \hspace{.4cm} 
	\overset{\mathfrak{su}_{m_0}}{2} \hspace{.2cm}
	\overset{\mathfrak{su}_{2m_0}}{2} \hspace{.2cm}
	\ldots
	\overset{\mathfrak{su}_{(\qappa-1)m_0}}{2} \, \,
	\underset{[N_f = m_0]}{\overset{\mathfrak{su}_{\qappa m_0}}{2}} \,\, 
	\underbrace{\overset{\mathfrak{su}_{\qappa m_0}}{2} \,\, \cdots \,\, \overset{\mathfrak{su}_{\qappa m_0}}{2}}_{N-\qappa-1} \,\,
	[{\rm SU}(\qappa m_0)].
\ee
These theories were studied in \cite{DelZotto:2014hpa, Zafrir:2015rga,Ohmori:2015tka, Bah:2017wxp}.
They can be called  the ``massive E-string theories'' as in the last reference, since they correspond to NS5-branes probing the O8-D8 combination in the presence of the Romans mass.  

The mirror of the $T^3$ compactification of \eref{genmassEstring} is
\be \label{mirgenmassEstring}
 {\Node{}{1}}-{\Node{}{2}}- \cdots -{\Node{}{\qappa m_0 }}-\node{}{N+r_1}-\node{}{2N+r_2}-\node{}{3N+r_{3}}-\node{}{4N+r_4}-\node{}{5N+r_5}-\node{\cver{}{3N+r_{3'}}}{6N+  r_6}-\node{}{4N+r_{4'}}-\node{}{2N+r_{2'}}
\ee
where the values of $r_i$ and the Kac labels for each $m_0$ are given in Table \ref{tab:m0ri}.  Note that
\be
\sum_i {r_i} = \frac{1}{2} \qappa m_0 (m_0-1)~.
\ee
The SCFT Higgs branch dimension of \eref{genmassEstring} is
\be \label{HiggsTE}
\dim^{\text{SCFT}}_\BH \text{Higgs of $\CT^{6d}_E(\qappa ,m_0, N)$} =  30N + \frac{1}{2} \qappa m_0^2(\qappa+1)  -1~;
\ee
this is equal to the Coulomb branch dimension of \eref{mirgenmassEstring}.

\subsection{Higgsing the $\SU(k)$ flavour symmetry}
In the theories we have discussed so far, there is always an $\SU(k)$ flavour symmetry which came from the gauge symmetry on the $\bC^2/\bZ_k$ singularity.
From the $3d$ quiver perspective, this symmetry arises from the topological symmetry associated with the nodes in the tail ${\Node{}{1}}-{\Node{}{2}}-\cdots -\Node{}{k}$. 

We can obtain another class of models by on nilpotent VEVs that Higgs the flavour symmetry $\SU(k)$.\footnote{The authors thank Alessandro Tomasiello for the discussion about this class of theories.}   Suppose that such VEVs are in the nilpotent orbit of $\SU(k)$ given by $\bigoplus_i J_{s_i}$ where $J_s$ is a $s\times s$ Jordan block so that $Y = [s_1, s_2, \ldots, s_\ell]$ is a corresponding partition of $k$.

Assuming that the 6d quiver theory before the Higgsing has a sufficiently long plateau of $\SU(k)$ gauge groups, this Higgsing can be performed exactly as in 4d class S theory e.g.~as described in Sec.~12.5 of \cite{Tachikawa:2013kta}. 
Its effect in 6d quiver was studied in \cite{Heckman:2016ssk,Mekareeya:2016yal}.
In the end, we see that the tail on the right-hand side of the quiver to have the form
\be \label{gluegenmassEstring1}
 \cdots \,\, \overset{\mathfrak{su}({k})}{2}
  \,\, \,\, 
  \underset{[N_f = u_{\ell'}]}{{\overset{\mathfrak{su}({k})}{2}}} \, \, \, \, \underset{[N_f = (u_{\ell'-1}-u_{\ell'})]}{{\overset{\mathfrak{su}{(k-u_{\ell'})}}{2}}} \, \, \, \,  \cdots 
  \, \, \, \, \underset{[N_f = (u_{2}-u_3)]}{\overset{\mathfrak{su}({u_2+u_1})}{2}}~
  \, \, \, \, \underset{[N_f = (u_{1}-u_2)]}{\overset{\mathfrak{su}({u_1})}{2}}~,
\ee
where $u_i$ are the elements of the transpose 
$ Y^T= [ u_1,  u_2, \ldots,  u_{\ell'}]$,
 and we define $u_i =0$ for $i > \ell'$. 

The SCFT Higgs branch dimension of \eref{gluegenmassEstring1} is
\be \label{HiggsdimgluegenmassEstring1}
 \dim^{\text{SCFT}}_\BH \text{Higgs of \eref{gluegenmassEstring1}}  
=   \left[30(\NN+k) - \langle \vec w, \vec \rho \rangle +\frac{1}{2} k(k+1) -1 \right] -\dim_\bH \cO_Y
\ee
where $\cO_Y$ is the nilpotent orbit labeled by $Y$.

The mirror of the $T^3$ compactification of \eref{gluegenmassEstring1} is
\be\label{mirrgenmassEstring1} 
\frac{T_{Y}(\SU(k))~ \times~ \sqnode{}{k} -\node{}{\tilde N_1}-\node{}{\tilde N_2}-\node{}{\tilde N_3}-\node{}{\tilde N_4}-\node{}{\tilde N_5}-\node{\cver{}{\tilde N_{3'}}}{\tilde N_6}-\node{}{\tilde N_{4'}}-\node{}{\tilde N_{2'}}}{\UU(k)/\UU(1)}.
\ee
 In other words, we simply replace the tail ${\Node{}{1}}-{\Node{}{2}}-\cdots -\sqNode{}{k}$ for the theories discussed in the preceding sections by $T_{Y}(\SU(k))$, where the latter is defined as in \cite{Gaiotto:2008ak}.   The Coulomb branch dimension of \eref{mirrgenmassEstring1} is
\be \label{Coul3dmirrgenmassEstring1}
\begin{split}
&\dim_\BH \text{Coulomb of \eref{mirrgenmassEstring1}} \\
&= \left[ 30(\NN+k) - \langle \vec w, \vec \rho \rangle \right] + \left[\frac{1}{2} \{ (k^2 -1) - (k-1) \}  -\dim_\bH\cO_Y  \right] + (k-1)\\
& = 30 (\NN+k) + \frac{1}{2}k(k+1) -1 - \dim_\bH \cO_Y - \langle \vec w, \vec \rho \rangle~,
\end{split}
\ee
where the terms in the second square brackets in the second line denote the Coulomb branch dimension of $T_{Y}(\SU(k))$.  This result is indeed in agreement with \eref{HiggsdimgluegenmassEstring1}. 

As an example, let us consider $\CT^{6d}_E(k ,{m_0 = 1}, N)$ of the previous section and perform the Higgsing with $Y=[k-1,1]$.  The resulting $6d$ theory is
\be
[E_8] \, \,  1  \, \,  \overset{\mathfrak{su}(1)}{2}  \, \,  \overset{\mathfrak{su}(2)}{2}  \,  \,\ldots  \, \,  \overset{\mathfrak{su}(k-1)}{2}  \, \, \underset{\left[N_f=1 \right]}{\overset{\mathfrak{su}(k)}{2}} \, \, \overset{\mathfrak{su}(k)^{N-2k}}{2}  \, \, \underset{\left[N_f=1 \right]}{\overset{\mathfrak{su}(k)}{2}}\, \, \overset{\mathfrak{su}(k-1)}{2}  \,\, \ldots \,\,  \overset{\mathfrak{su}(2)}{2} \, \,  \overset{\mathfrak{su}(1)}{2}~,
\ee
where the number of tensor multiplets is $N$.  This theory is similar to that discussed in (36) of \cite{Aspinwall:1997ye}, (5.2) of \cite{Hanany:1997gh}, except that we have only one $(-1)$-curve in the quiver, instead of two.  The mirror of the $T^3$ compactification of this theory is
\be
{\Node{}{1}} -{\Node{}{k}} -\node{}{N}-\node{}{2N}-\node{}{3N}-\node{}{4N}-\node{}{5N}-\node{\cver{}{3N}}{6N}-\node{}{4N}-\node{}{2N}~, \label{E8E8quiv}
\ee
This quiver is a ``good'' theory in the sense of \cite{Gaiotto:2008ak} if $N+1\geq2k$ and $k \geq 2$. In this case, this quiver is the $3d$ mirror theory of the $S^1$ reduction of the class $\CS$ theory of type $\SU(6N)$ associated a sphere with the punctures
\be
[N^5, N-k,k-1,1], \quad [(3N)^2], \quad [(2N)^3]~.
\ee

\section*{Acknowledgments}
The authors thank  Hiroyuki Shimizu for the collaboration at the early stages. 
NM sincerely thanks Stefano Cremonesi, Amihay Hanany and Alessandro Tomasiello for a close collaboration, invaluable insights, and several useful discussions. He also grateful to the hospitality of the organisers of the Pollica Summer Workshop 2017, including Fernando Alday, Philip Argyres, Madalena Lemos and Mario Martone.  He is supported in part by the INFN, the ERC Starting Grant 637844- HBQFTNCER, as well as the ERC STG grant 306260 through the Pollica Summer Workshop.
KO  gratefully acknowledges support from the Institute for Advanced Study.
YT is partially supported in part byJSPS KAKENHI Grant-in-Aid (Wakate-A), No.17H04837 
and JSPS KAKENHI Grant-in-Aid (Kiban-S), No.16H06335.
YT and GZ are partially supported by WPI Initiative, MEXT, Japan at IPMU, the University of Tokyo.

\bibliographystyle{ytphys}
\bibliography{refs}

\providecommand{\href}[2]{#2}\begingroup\raggedright\begin{thebibliography}{10}

\bibitem{Atiyah:1978ri}
M.~F. Atiyah, N.~J. Hitchin, V.~G. Drinfeld, and Y.~I. Manin, ``{Construction
  of Instantons},''
\href{http://dx.doi.org/10.1016/0375-9601(78)90141-X}{{\em Phys. Lett.}
  {\bfseries A65} (1978) 185--187}.

\bibitem{KronheimerNakajima}
P.~B. Kronheimer and H.~Nakajima, ``Yang-{M}ills instantons on {ALE}
  gravitational instantons,'' \href{http://dx.doi.org/10.1007/BF01444534}{{\em
  Math. Ann.} {\bfseries 288} no.~2, (1990) 263--307}.

\bibitem{Bianchi:1996zj}
M.~Bianchi, F.~Fucito, G.~Rossi, and M.~Martellini, ``{Explicit Construction of
  Yang-Mills Instantons on ALE Spaces},''
  \href{http://dx.doi.org/10.1016/0550-3213(96)00240-4}{{\em Nucl.Phys.}
  {\bfseries B473} (1996) 367--404},
\href{http://arxiv.org/abs/hep-th/9601162}{{\ttfamily arXiv:hep-th/9601162}}.

\bibitem{Witten:1995gx}
E.~Witten, ``{Small Instantons in String Theory},''
  \href{http://dx.doi.org/10.1016/0550-3213(95)00625-7}{{\em Nucl. Phys.}
  {\bfseries B460} (1996) 541--559},
\href{http://arxiv.org/abs/hep-th/9511030}{{\ttfamily arXiv:hep-th/9511030}}.

\bibitem{Douglas:1996sw}
M.~R. Douglas and G.~W. Moore, ``{D-Branes, Quivers, and ALE Instantons},''
\href{http://arxiv.org/abs/hep-th/9603167}{{\ttfamily arXiv:hep-th/9603167}}.

\bibitem{Aspinwall:1997ye}
P.~S. Aspinwall and D.~R. Morrison, ``{Point-Like Instantons on K3
  Orbifolds},'' \href{http://dx.doi.org/10.1016/S0550-3213(97)00516-6}{{\em
  Nucl. Phys.} {\bfseries B503} (1997) 533--564},
\href{http://arxiv.org/abs/hep-th/9705104}{{\ttfamily arXiv:hep-th/9705104}}.

\bibitem{Gaiotto:2009we}
D.~Gaiotto, ``{${\mathcal{N}}\!=2$ Dualities},''
  \href{http://dx.doi.org/10.1007/JHEP08(2012)034}{{\em JHEP} {\bfseries 1208}
  (2012) 034},
\href{http://arxiv.org/abs/0904.2715}{{\ttfamily arXiv:0904.2715 [hep-th]}}.

\bibitem{Cremonesi:2013lqa}
S.~Cremonesi, A.~Hanany, and A.~Zaffaroni, ``{Monopole Operators and Hilbert
  Series of Coulomb Branches of 3D ${\mathcal{N}}\!=4$ Gauge Theories},''
\href{http://arxiv.org/abs/1309.2657}{{\ttfamily arXiv:1309.2657 [hep-th]}}.

\bibitem{Heckman:2013pva}
J.~J. Heckman, D.~R. Morrison, and C.~Vafa, ``{On the Classification of 6D
  SCFTs and Generalized ADE Orbifolds},''
\href{http://arxiv.org/abs/1312.5746}{{\ttfamily arXiv:1312.5746 [hep-th]}}.

\bibitem{DelZotto:2014hpa}
M.~Del~Zotto, J.~J. Heckman, A.~Tomasiello, and C.~Vafa, ``{6D Conformal
  Matter},''
\href{http://arxiv.org/abs/1407.6359}{{\ttfamily arXiv:1407.6359 [hep-th]}}.

\bibitem{Heckman:2015bfa}
J.~J. Heckman, D.~R. Morrison, T.~Rudelius, and C.~Vafa, ``{Atomic
  Classification of 6D SCFTs},''
\href{http://arxiv.org/abs/1502.05405}{{\ttfamily arXiv:1502.05405 [hep-th]}}.

\bibitem{Zafrir:2015rga}
G.~Zafrir, ``{Brane Webs, $5d$ Gauge Theories and $6d$ $\mathcal{N}=(1,0)$
  SCFT's},'' \href{http://dx.doi.org/10.1007/JHEP12(2015)157}{{\em JHEP}
  {\bfseries 12} (2015) 157},
\href{http://arxiv.org/abs/1509.02016}{{\ttfamily arXiv:1509.02016 [hep-th]}}.

\bibitem{Ohmori:2015tka}
K.~Ohmori and H.~Shimizu, ``{$S^1/T^2$ Compactifications of 6D $
  \mathcal{N}=(1,0) $ Theories and Brane Webs},''
  \href{http://dx.doi.org/10.1007/JHEP03(2016)024}{{\em JHEP} {\bfseries 03}
  (2016) 024},
\href{http://arxiv.org/abs/1509.03195}{{\ttfamily arXiv:1509.03195 [hep-th]}}.

\bibitem{Hayashi:2015zka}
H.~Hayashi, S.-S. Kim, K.~Lee, and F.~Yagi, ``{6D SCFTs, 5D Dualities and Tao
  Web Diagrams},''
\href{http://arxiv.org/abs/1509.03300}{{\ttfamily arXiv:1509.03300 [hep-th]}}.

\bibitem{Ohmori:2015pua}
K.~Ohmori, H.~Shimizu, Y.~Tachikawa, and K.~Yonekura, ``{6D $\mathcal{N}=(1,0)$
  Theories on $T^2$ and Class S Theories: Part I},''
  \href{http://dx.doi.org/10.1007/JHEP07(2015)014}{{\em JHEP} {\bfseries 07}
  (2015) 014},
\href{http://arxiv.org/abs/1503.06217}{{\ttfamily arXiv:1503.06217 [hep-th]}}.

\bibitem{KronheimerNilpotent}
P.~B. Kronheimer, ``Instantons and the geometry of the nilpotent variety,''
  \href{http://projecteuclid.org/euclid.jdg/1214445316}{{\em J. Differential
  Geom.} {\bfseries 32} no.~2, (1990) 473--490}.

\bibitem{Tachikawa:2014qaa}
Y.~Tachikawa, ``{Moduli Spaces of S$O(8)$ Instantons on Smooth ALE Spaces as
  Higgs Branches of 4D ${\mathcal{N}}\!=2$ Supersymmetric Theories},''
  \href{http://dx.doi.org/10.1007/JHEP06(2014)056}{{\em JHEP} {\bfseries 06}
  (2014) 056},
\href{http://arxiv.org/abs/1402.4200}{{\ttfamily arXiv:1402.4200 [hep-th]}}.

\bibitem{Benini:2010uu}
F.~Benini, Y.~Tachikawa, and D.~Xie, ``{Mirrors of 3D Sicilian Theories},''
  \href{http://dx.doi.org/10.1007/JHEP09(2010)063}{{\em JHEP} {\bfseries 09}
  (2010) 063},
\href{http://arxiv.org/abs/1007.0992}{{\ttfamily arXiv:1007.0992 [hep-th]}}.

\bibitem{Nakajima:2015txa}
H.~Nakajima, ``{Towards a Mathematical Definition of Coulomb Branches of
  $3$-dimensional $\mathcal{N}=4$ Gauge Theories, I},''
  \href{http://dx.doi.org/10.4310/ATMP.2016.v20.n3.a4}{{\em Adv. Theor. Math.
  Phys.} {\bfseries 20} (2016) 595--669},
\href{http://arxiv.org/abs/1503.03676}{{\ttfamily arXiv:1503.03676 [math-ph]}}.

\bibitem{Nakajima:2015gxa}
H.~Nakajima, ``{Questions on Provisional Coulomb Branches of $3$-dimensional
  ${\mathcal{N}}\!=4$ Gauge Theories},''
\href{http://arxiv.org/abs/1510.03908}{{\ttfamily arXiv:1510.03908 [math-ph]}}.

\bibitem{Cremonesi:2014xha}
S.~Cremonesi, G.~Ferlito, A.~Hanany, and N.~Mekareeya, ``{Coulomb Branch and
  the Moduli Space of Instantons},''
  \href{http://dx.doi.org/10.1007/JHEP12(2014)103}{{\em JHEP} {\bfseries 12}
  (2014) 103},
\href{http://arxiv.org/abs/1408.6835}{{\ttfamily arXiv:1408.6835 [hep-th]}}.

\bibitem{Mekareeya:2015bla}
N.~Mekareeya, ``{The Moduli Space of Instantons on an ALE Space from 3D
  $\mathcal{N}=4$ Field Theories},''
  \href{http://dx.doi.org/10.1007/JHEP12(2015)174}{{\em JHEP} {\bfseries 12}
  (2015) 174},
\href{http://arxiv.org/abs/1508.06813}{{\ttfamily arXiv:1508.06813 [hep-th]}}.

\bibitem{Kac}
V.~G. Kac, {\em {Infinite Dimensional Lie Algebras}}.
\newblock {Cambridge University Press}, 1994.

\bibitem{Nakajima}
H.~Nakajima, ``Moduli spaces of anti-self-dual connections on {ALE}
  gravitational instantons,'' \href{http://dx.doi.org/10.1007/BF01233429}{{\em
  Invent. Math.} {\bfseries 102} no.~2, (1990) 267--303}.

\bibitem{Ohmori:2014kda}
K.~Ohmori, H.~Shimizu, Y.~Tachikawa, and K.~Yonekura, ``{Anomaly Polynomial of
  General 6D SCFTs},'' \href{http://dx.doi.org/10.1093/ptep/ptu140}{{\em PTEP}
  {\bfseries 2014} no.~10, (2014) 103B07},
\href{http://arxiv.org/abs/1408.5572}{{\ttfamily arXiv:1408.5572 [hep-th]}}.

\bibitem{Intriligator:2014eaa}
K.~Intriligator, ``{6D, $\mathcal{N}=(1,0)$ Coulomb Branch Anomaly Matching},''
\href{http://arxiv.org/abs/1408.6745}{{\ttfamily arXiv:1408.6745 [hep-th]}}.

\bibitem{Ohmori:2014pca}
K.~Ohmori, H.~Shimizu, and Y.~Tachikawa, ``{Anomaly Polynomial of E-String
  Theories},'' \href{http://dx.doi.org/10.1007/JHEP08(2014)002}{{\em JHEP}
  {\bfseries 08} (2014) 002},
\href{http://arxiv.org/abs/1404.3887}{{\ttfamily arXiv:1404.3887 [hep-th]}}.

\bibitem{Brunner:1997gf}
I.~Brunner and A.~Karch, ``{Branes at Orbifolds Versus Hanany Witten in
  Six-Dimensions},''
  \href{http://dx.doi.org/10.1088/1126-6708/1998/03/003}{{\em JHEP} {\bfseries
  9803} (1998) 003},
\href{http://arxiv.org/abs/hep-th/9712143}{{\ttfamily arXiv:hep-th/9712143}}.

\bibitem{Hanany:1997gh}
A.~Hanany and A.~Zaffaroni, ``{Branes and six-dimensional supersymmetric
  theories},'' \href{http://dx.doi.org/10.1016/S0550-3213(98)00355-1}{{\em
  Nucl. Phys.} {\bfseries B529} (1998) 180--206},
\href{http://arxiv.org/abs/hep-th/9712145}{{\ttfamily arXiv:hep-th/9712145}}.

\bibitem{BergmanGomez}
O.~Bergman and D.~R{odr\'\i guez-G\'omez}, ``{5d quivers and their AdS$_6$
  duals},'' \href{http://dx.doi.org/10.1007/JHEP07(2012)171}{{\em JHEP}
  {\bfseries 07} (2012) 171}, \href{http://arxiv.org/abs/1206.3503}{{\ttfamily
  arXiv:1206.3503 [hep-th]}}.

\bibitem{HayashiKimLee}
H.~Hayashi, S.-S. Kim, K.~Lee, M.~Taki, and F.~Yagi, ``{A new 5d description of
  6d D-type minimal conformal matter},''
  \href{http://dx.doi.org/10.1007/JHEP08(2015)097}{{\em JHEP} {\bfseries 08}
  (2015) 097}, \href{http://arxiv.org/abs/1505.04439}{{\ttfamily
  arXiv:1505.04439 [hep-th]}}.

\bibitem{HayashiKimLee2}
H.~Hayashi, S.-S. Kim, K.~Lee, M.~Taki, and F.~Yagi, ``{More on 5d descriptions
  of 6d SCFTs},'' \href{http://dx.doi.org/10.1007/JHEP10(2016)126}{{\em JHEP}
  {\bfseries 10} (2016) 126}, \href{http://arxiv.org/abs/1512.08239}{{\ttfamily
  arXiv:1512.08239 [hep-th]}}.

\bibitem{BergmanZafrir}
O.~Bergman and G.~Zafrir, ``{5d fixed points from brane webs and O7-planes},''
  \href{http://dx.doi.org/10.1007/JHEP12(2015)163}{{\em JHEP} {\bfseries 12}
  (2015) 163}, \href{http://arxiv.org/abs/1507.03860}{{\ttfamily
  arXiv:1507.03860 [hep-th]}}.

\bibitem{BBT}
F.~Benini, S.~Benvenuti, and Y.~Tachikawa, ``{Webs of five-branes and N=2
  superconformal field theories},''
  \href{http://dx.doi.org/10.1088/1126-6708/2009/09/52}{{\em JHEP} {\bfseries
  09} (2009) 052}, \href{http://arxiv.org/abs/0906.0359}{{\ttfamily
  arXiv:0906.0359 [hep-th]}}.

\bibitem{Chacaltana:2010ks}
O.~Chacaltana and J.~Distler, ``{Tinkertoys for Gaiotto Duality},''
  \href{http://dx.doi.org/10.1007/JHEP11(2010)099}{{\em JHEP} {\bfseries 1011}
  (2010) 099},
\href{http://arxiv.org/abs/1008.5203}{{\ttfamily arXiv:1008.5203 [hep-th]}}.

\bibitem{Tachikawa:2015bga}
Y.~Tachikawa, ``{A Review of the $T_N$ Theory and Its Cousins},''
\href{http://arxiv.org/abs/1504.01481}{{\ttfamily arXiv:1504.01481 [hep-th]}}.

\bibitem{Gaiotto:2008ak}
D.~Gaiotto and E.~Witten, ``{S-Duality of Boundary Conditions in
  ${\mathcal{N}}\!=4$ Super Yang-Mills Theory},''
  \href{http://dx.doi.org/10.4310/ATMP.2009.v13.n3.a5}{{\em
  Adv.Theor.Math.Phys.} {\bfseries 13} (2009) 721},
\href{http://arxiv.org/abs/0807.3720}{{\ttfamily arXiv:0807.3720 [hep-th]}}.

\bibitem{Bullimore:2015lsa}
M.~Bullimore, T.~Dimofte, and D.~Gaiotto, ``{The Coulomb Branch of 3D
  $\mathcal{N}=4$ Theories},''
  \href{http://dx.doi.org/10.1007/s00220-017-2903-0}{{\em Commun. Math. Phys.}
  {\bfseries 354} no.~2, (2017) 671--751},
\href{http://arxiv.org/abs/1503.04817}{{\ttfamily arXiv:1503.04817 [hep-th]}}.

\bibitem{Ohmori:2015pia}
K.~Ohmori, H.~Shimizu, Y.~Tachikawa, and K.~Yonekura, ``{6D
  $\mathcal{N}=\left(1,\;0\right) $ Theories on $S^1/T^2$ and Class S Theories:
  Part II},'' \href{http://dx.doi.org/10.1007/JHEP12(2015)131}{{\em JHEP}
  {\bfseries 12} (2015) 131},
\href{http://arxiv.org/abs/1508.00915}{{\ttfamily arXiv:1508.00915 [hep-th]}}.

\bibitem{Gaiotto:2012uq}
D.~Gaiotto and S.~S. Razamat, ``{Exceptional Indices},''
  \href{http://dx.doi.org/10.1007/JHEP05(2012)145}{{\em JHEP} {\bfseries 05}
  (2012) 145},
\href{http://arxiv.org/abs/1203.5517}{{\ttfamily arXiv:1203.5517 [hep-th]}}.

\bibitem{Cremonesi:2014vla}
S.~Cremonesi, A.~Hanany, N.~Mekareeya, and A.~Zaffaroni, ``{Coulomb branch
  Hilbert series and Three Dimensional Sicilian Theories},''
  \href{http://dx.doi.org/10.1007/JHEP09(2014)185}{{\em JHEP} {\bfseries 09}
  (2014) 185},
\href{http://arxiv.org/abs/1403.2384}{{\ttfamily arXiv:1403.2384 [hep-th]}}.

\bibitem{Bah:2017wxp}
I.~Bah, A.~Passias, and A.~Tomasiello, ``{AdS5 compactifications with punctures
  in massive IIA supergravity},''
\href{http://arxiv.org/abs/1704.07389}{{\ttfamily arXiv:1704.07389 [hep-th]}}.

\bibitem{Tachikawa:2013kta}
Y.~Tachikawa, ``{${\mathcal{N}}\!=2$ Supersymmetric Dynamics for
  Pedestrians},'' \href{http://dx.doi.org/10.1007/978-3-319-08822-8}{{\em
  Lect.Notes Phys.} {\bfseries 890} (2013) 2014},
\href{http://arxiv.org/abs/1312.2684}{{\ttfamily arXiv:1312.2684 [hep-th]}}.

\bibitem{Heckman:2016ssk}
J.~J. Heckman, T.~Rudelius, and A.~Tomasiello, ``{6D RG Flows and Nilpotent
  Hierarchies},'' \href{http://dx.doi.org/10.1007/JHEP07(2016)082}{{\em JHEP}
  {\bfseries 07} (2016) 082},
\href{http://arxiv.org/abs/1601.04078}{{\ttfamily arXiv:1601.04078 [hep-th]}}.

\bibitem{Mekareeya:2016yal}
N.~Mekareeya, T.~Rudelius, and A.~Tomasiello, ``{T-branes, Anomalies and Moduli
  Spaces in 6D SCFTs},''
\href{http://arxiv.org/abs/1612.06399}{{\ttfamily arXiv:1612.06399 [hep-th]}}.

\end{thebibliography}\endgroup

\end{document}